\documentstyle[aps,pra,preprint,floats,epsfig,array]{revtex}

\tightenlines

\begin{document}

\hfuzz 2pt \vfuzz 2pt %\draft

\def\Journal#1#2#3#4{{\em #1}, {\bf #2}, #3 (#4)}

%-----------------------------------------------------
\title{\Large\bf 
	Spatial overlap of nonclassical ultrashort light pulses and
	formation of polarization-squeezed light
	}

\author{
	F. Popescu\footnote{E-mail:~florentin\_p@hotmail.com}
	}

\address{
	Physics Department, 
	Florida State University, 
	Tallahassee,
        Florida, 32306
	}

\date{August 19, 2004}

\maketitle

\begin{abstract} {\small
{ We investigate the spatial overlap
of nonclassical ultrashort light pulses produced by self-phase
modulation effect in electronic Kerr media and its relevance in the
formation of polarization-squeezed states of light. The light
polarization is treated in terms of four quantum Stokes parameters
whose spectra of quantum fluctuations are investigated. We show that
the frequency at which the suppression of quantum fluctuations of
Stokes parameters is the greatest can be controlled by adjusting the
linear phase difference between pulses. By varying the intensity of
one pulse one can suppress effectively the quantum fluctuations of
Stokes parameters. We study the overlap of nonclassical pulses
inside of an anisotropic electronic Kerr medium and we show that the
cross-phase modulation effect can be employed to control the
polarization-squeezed state of light. Moreover, we establish that
the change of the intensity or of the nonlinear phase shift per
photon for one pulse controls effectively the squeezing of Stokes
parameters. The spatial overlap of a coherent pulse field with an
interference pulse produced by mixing two quadrature-squeezed pulses
on a beam splitter is analyzed. It is found that squeezing is
produced in three of the four Stokes parameters, with the squeezing
in the first two being simultaneous. }
}
\end{abstract}

\vspace{5mm}


\vfill

%---------------------------------------------------------------------
\section{Introduction}\label{Introduction}
Quantum states of light with squeezed polarization have recently
been in the focus of theoretical and experimental investigations.
The quantum analysis of the light polarization is generally based on
four Stokes parameters associated with four Hermitian Stokes
operators \cite{Agarwal,Tanas}. By definition, the quantum state of
light with the level of quantum fluctuations of Stokes operators
smaller than the level corresponding to the coherent state is called
polarization-squeezed (PS) state. Its existence was predicted
theoretically in \cite{Orlov} and was revealed experimentally in a
series of recent experiments \cite{Hald,Heersnik,Heersnik2}.
Historically, the first experimental realization of a PS state of
light implied the mixing of a strong orthogonally-polarized coherent
beam with squeezed vacuum on a 50/50 beam splitter (BS) \cite{Hald}.
The optical parametric amplification ($\chi^{2}$) has also been
successfully employed to generate the PS state of light
\cite{Grangier,Bowen1,Bowen2}. In the last decade the improvement of
fiber optics techniques in the pulse field regime facilitated
experiments directed to produce nonclassical solitons
\cite{Rosenbluh,Margalit} or highly stable zero-dispersion
quadrature-squeezed ultrashort light pulses (USPs)
\cite{Bergman1,Bergman2,Bergman3}. Recently it has been found
experimentally that the simple spatial overlap of two orthogonally
polarized quadrature-squeezed pulses leads to the formation of the
PS light \cite{Heersnik,Heersnik2}.

The self-phase modulation (SPM) effect, which is responsible for the
quadrature squeezing of pulses in optical Keer-like media
($\chi^{3}$), does not influence the photon statistics of pulses
since the photon number is a constant of motion \cite{Kitagawa}. It
was noted for the first time in \cite{Blow}, that a consistent
time-dependent quantum description of the SPM effect must
necessarily account for the additional noise related to nonlinear
absorbtion (the imaginary part of nonlinear susceptibility). In
\cite{Boivin} the Kerr nonlinearity has been treated as a Raman-like
one, the quantum and thermal noises being considered as a
fluctuating addition to the relaxation nonlinearity in the
interaction Hamiltonian. The addition is necessary in order to
preserve the commutation relations among the pulse field-amplitude
operators.  Hence, the resulting quantum equation of motion for the
total field includes the Hermitian-phase noise operator which
describes the coupling of the field to a collection of localized,
independent, medium oscillators. Besides, the average value of the
noise operator on the coherent state is considered to be zero. The
time-delayed Raman response of nonlinearity, which is around $50$ fs
in fused-silica fibers, varies over frequencies of interest and is
caused by back action of nonlinear nuclear vibrations on electronic
ones. However, the contribution of Raman oscillators to the Kerr
effect is secondary to the one of the electronic motion. Indeed, if
we deal with the nonlinear propagation of USPs, for instance through
fused-silica fibers, since the Raman oscillators contribute with
less than $20\%$ to the Kerr effect, the contribution of the
electronic motion on $\sim1$ fs time scale is more than $80\%$
\cite{Joneckis}. The analysis of the SPM effect then has to focus on
contribution of the electronic motion, especially in the case of
USPs with a duration much less than the time-delayed Raman response
or in case the pulse's frequencies are far from any Raman
resonances. An attempt to develop a quantum theory of the pulse SPM
primarily due to the electronic motion has been undertaken in
\cite{Boivin1}, but the electronic nonlinearity has been modeled as
a Raman-like one. The electronic nonlinear response function in the
frequency domain over the bandwidths of interest is considered
constant and approximated with a delta function in the time domain.
This implies an instantaneous electronic response of the Kerr
nonlinearity. In real situations the nonlinear response of
electronic nonlinearity is finite. It results in a non-delta
relaxation function in the proper analytical description of the Kerr
effect. Such a consistent description based on the momentum
operator, which is connected with the evolution of the field in
space and incorporates the relaxation function of the electronic
nonlinearity, has been implemented in \cite{POP99,POP00}, where the
correlation functions and spectra of quadrature components of USPs
subjected to the SPM effect in the electronic Kerr medium are
investigated. Since the momentum operator is in the normally ordered
form, in order to satisfy the commutation relations for
time-dependent Bose operators there is no need for additional
fluctuating terms. The approach developed in \cite{POP99,POP00} for
the SPM effect has been extended in \cite{QSO} to the case of two
USPs undergoing, besides the SPM effect, the cross-phase modulation
(XPM) effect.

In this paper we apply the quantum theory of the SPM and combined
SPM-XPM effects developed in \cite{POP99,POP00} and \cite{QSO},
respectively, to the case of spatial overlap and interference on a
BS of quadrature-squeezed pulses. The nonclassical USPs are obtained
by undergoing the SPM effect in an inertial electronic Kerr medium.
The finite relaxation time of the electronic Kerr nonlinearity is
accounted for and the dispersion of linear properties is described
in the first approximation of the dispersion theory. We show that
the approach for the SPM effect based on the momentum operator
allows an adequate calculation of correlation functions and
fluctuation spectra of quantum Stokes operators.

The paper is organized as follows. In Section \ref{Section1} the
quantum model of SPM effect for USPs based on the momentum operator
for the pulse field is exposed briefly. We present the
space-evolution equation for time dependent Bose operators, its
solution, as well as some elements of the algebra of Bose operators.
Section \ref{Section2} introduces the quantum characterization of
polarization of light in terms of Stokes operators, as well as their
correlation functions and spectra. In Section \ref{Section3} we
analyze the overlap between a coherent USP and a quadrature-squeezed
one, the latter USP being produced by employing the SPM effect in an
electronic isotropic Kerr medium. The average values, the
correlation functions of Stokes operators $\hat{S}_{2}$ and
$\hat{S}_{3}$ are computed, and the spectrum of $\hat{S}_{2}$ is
graphically represented. This analysis is continued in Section
\ref{Section4} by investigating the spatial overlap of two
independent quadrature-squeezed USPs. In Section \ref{Section5} we
study the overlap of two quadrature-squeezed pulses inside a
nonlinear anisotropic Kerr medium in the presence of the combined
SPM-XPM effect. The momentum operator for the XPM effect introduces
the section. Here we show that the XPM effect can be employed to
control the formation of polarization-squeezed spectra. Section
\ref{Section6} is dedicated to the interference on a BS of two
independent quadrature-squeezed USPs and their overlap with a
coherent USP or with squeezed vacuum. We reveal that the squeezing
is possible in three Stokes parameters $\hat{S}_{0}$, $\hat{S}_{1}$,
and  $\hat{S}_{2}$. We compare our analytical results with the
experimental ones reported in \cite{Heersnik,Heersnik2}. Our
concluding remarks are presented in Section \ref{Conclusion}.
%----------------------------------------------------------------------------------
\section{SPM effect in electronic Kerr medium}\label{Section1}
The traditional way to derive the quantum equation for the SPM
effect is based on the interaction Hamiltonian. As a consequence one
obtains a time-evolution equation. The transition to space-evolution
equation is usually realized by enforcing the replacement
$t\rightarrow z/u$, where $z$ is the distance passed inside the Kerr
medium and $u$ is the group velocity. This approach is reasonable
for single-mode radiation. However, the analytical description of
the nonlinear pulse propagation contains both $t$ and $z$. The
consistent quantum description of the Kerr effect requires then the
use of the momentum operator connected with the evolution of the
pulse field in space \cite{Mooki}.

When accounting for the finite relaxation time of the Kerr
nonlinearity the SPM effect is described with the following momentum
operator \cite{POP99,POP00,QSO}:
\begin{equation}
\hat{G}_{\rm{spm}}(z)=\hbar\beta\int_{-\infty}^{\infty}dt
\int_{-\infty}^{t}H(t-t_{1})\hat{\mathbf{N}}\bigl
[\hat{n}(t,z)\hat{n}(t_{1},z)\bigl]dt_{1}\label{spm}
\end{equation}
where $\hbar$ is the Plank constant, the factor $\beta$ is defined
by the Kerr nonlinearity of the medium
($\beta=\hbar\omega_{0}k_{0}n_{2}/8n_{0}V$ \cite{QSO,Ahmanov}),
$\hat{\mathbf{N}}$ is the normal ordering operator, $H(t)$ is the
causal nonlinear response function [$H(t)\neq 0$ at $t\geq 0$ and
$H(t)=0$ at $t<0$], $\hat{n}(t,z)=\hat{A}^{+}(t,z)\hat{A}(t,z)$ is
the ``photon number density" operator in the cross-section $z$ of
the medium, and $\hat{A}(t,z)$ and $\hat{A}^{+}(t,z)$ are the photon
annihilation and creation Bose operators with the commutation
relation
$[\hat{A}(t_{1},z),\hat{A}^{+}(t_{2},z)]=\delta{(t_{2}-t_{1})}$,
respectively. The thermal noise is neglected and the expression
(\ref{spm}) is averaged over the thermal fluctuations. In our
approach the pulse duration $\tau_{p}$ is much greater than the
relaxation time $\tau_{r}$, and the Kerr medium is lossless and
dispersionless; that is, the frequencies of the pulse are off
resonance.

In Heisenberg representation, the space evolution of $\hat{A}(t,z)$
is given by the equation \cite{Mooki}
\begin{equation}\label{evolution}
i\hbar\frac{\partial\hat{A}(t,z)}{\partial
z}=\left[\hat{G}_{\rm{spm}}(z),\hat{A}(t,z)\right].
\end{equation}
Then, with (\ref{spm}) the space evolution equation of
$\hat{A}(t,z)$ in the moving coordinate system ($z=z^{'}$,
$t=t^{'}-z/u$, where $t^{'}$ is the running time and $u$ is the
group velocity)
\begin{equation}\label{part1}
\frac{\partial\hat{A}(t,z)}{\partial z}-\hat{O}(t)\hat{A}(t,z)=0
\end{equation}
has the solution given by
\begin{equation}\label{anih}
\hat{A}(t,z)=e^{\hat{O}(t)}\hat{A}(t),
\end{equation}
where $\hat{O}(t)=i\gamma q(\hat{n}(t))$, $\gamma=\beta z$,
$\hat{n}(t)=\hat{A}^{+}(t)\hat{A}(t)$ is the ``photon number
density" operator at the entrance into the medium,
$\hat{n}(t)=\hat{n}(t,z=0)$, and
$q(\hat{n}(t))=\int_{-\infty}^{\infty}h(t_{1})\hat{n}(t-t_{1})dt_{1}$,
with $h(t)=H(|t|)$.

In comparison with the so-called nonlinear Schr\"{o}dinger equation,
used in the quantum theory of optical solitons (see for instance
\cite{Konig} and references therein), in (\ref{part1}) the light
pulse dispersion spreading in the medium is absent. This approach
corresponds to the first approximation of the dispersion theory
\cite{Ahmanov}. Note that the structure of $q(\hat{n}(t))$ is like
that of the linear response in the quantum description of a USP
spreading in the second-order approximation.

The statistical features of pulses at the output of the medium can
be evaluated by using the algebra of time-dependent Bose operators
\cite{POP99,POP00}. For example, in such algebra we have a
permutation relation of the form:
\begin{eqnarray}
\hat{A}(t_{1})e^{\hat{O}(t_{2})}=e^{\hat{O}(t_{2})
+{\mathcal{D}}(t_{2}-t_{1})}\hat{A}(t_{1}) \label{permut}
\end{eqnarray}
where ${\mathcal{D}}(t_{2}-t_{1})=i\gamma h(t_{2}-t_{1})$, $2\gamma$
is the nonlinear phase shift per photon. By using the theorem of
normal ordering \cite{POP99,POP00} one obtains the average values of
the Bose operators over the initial coherent summary state
$|\alpha(t)\rangle$:
\begin{eqnarray}
\langle e^{\hat{O}(t)}\rangle &=& e^{i\phi(t)-\mu(t)},\label{average1}\\
\langle e^{\hat{O}(t_{1})+\hat{O}(t_{2})}\rangle &=&
e^{i\phi(t_{1})+i\phi(t_{2})-\mu(t_{1})-\mu(t_{2})-
{\mathcal{K}}(t_{1},t_{2})}.\label{average2}
\end{eqnarray}
The parameters $\phi(t)=2\gamma\bar{n}(t)$,
$\mu(t)=\gamma^{2}\bar{n}(t)/2$ are connected with SPM of the pulse.
Here $\phi(t)$ is the nonlinear phase addition caused by SPM. The
eigenvalue $\alpha(t)$ of the annihilation operator $\hat{A}(t)$
over the coherent state $|\alpha(t)\rangle$ can be written as
$\alpha(t)=|\alpha(t)| e^{i\varphi(t)}$, where $\varphi(t)$ is the
linear phase of the pulse. Then
$|\alpha(t)|^{2}=\langle\hat{n}(t)\rangle\equiv\bar{n}(t)$. The time
dependence can be separated in $|\alpha(t)|$ by introducing the
pulse's envelope $r(t)$ so that $|\alpha(t)|=|\alpha(0)|r(t)$ with
$r(0)=1$. For the simplicity of notations let
$\phi(t=0)\equiv\phi_{0}$ and $\bar{n}(t=0)\equiv\bar{n}_{0}$. In
(\ref{average1}) and (\ref{average2})
${\mathcal{K}}(t_{1},t_{2})=\mu_{0}\,r^{2}(t_{1}+\tau/2)g(\tau)$ is
the temporal correlator due to SPM, where
$\mu_{0}=\gamma\bar{n}_{0}$, $g(\tau)=(1+|\tau|/\tau_{r})h(\tau)$,
and $\tau=t_{2}-t_{1}$.

Notice that in our approach there is no assumption about the form of
the relaxation function $H(t)$ of the Kerr nonlinearity. It is
positive at any instant $t>0$ and zero otherwise. If the SPM efect
is due primarily to the electronic motion occurring on $\sim 1$ fs
time scale ($\tau_{r}\leq 1$ fs), then in the absence of one- and
two-photon and Raman resonances, the relaxation function at $t\geq
0$ can be approximated in the form \cite{Ahmanov}:
\begin{equation}\label{relaxation}
H(t)=(1/\tau_r)\exp{\left(-t/\tau_r\right)}.
\end{equation}
In the following sections this specific form of the relaxation
function is used when evaluating the Fourier transform of $h(\tau)$
and $g(\tau)$.
%------------------------------------------------------------------
\section{Quantum characterization of the light polarization; Stokes
operators}\label{Section2}

The quantum description of the polarization of light is usually done
in terms of four Stokes operators $\hat{S}_{\alpha}$
($\alpha=\overline{0,3\mathstrut}$) \cite{Agarwal,Tanas}. For two
spatially overlapping USPs the Stokes operators are introduced as
follows:
\begin{eqnarray}
  \hat{S}_{0}(t) &=& \hat{A}^{+}_{1}(t)\hat{A}_{1}(t)+\hat{A}^{+}_{2}(t)\hat{A}_{2}(t)~~\,
  =Tr[\hat{D}^{+}_{1}(t)\hat{D}_{1}(t)],\label{S0}\\
  \hat{S}_{1}(t) &=& \hat{A}^{+}_{1}(t)\hat{A}_{1}(t)-\hat{A}^{+}_{2}(t)\hat{A}_{2}(t)~~\,
  =Tr[\hat{\sigma}_{3}\hat{D}^{+}_{1}(t)\hat{D}_{1}(t)],\label{S1}\\
  \hat{S}_{2}(t) &=& \hat{A}^{+}_{2}(t)\hat{A}_{1}(t)+\hat{A}^{+}_{1}(t)\hat{A}_{2}(t)~~\,
  =Tr[\hat{\sigma}_{1}\hat{D}^{+}_{2}(t)\hat{D}_{1}(t)],\label{S2}\\
  \hat{S}_{3}(t) &=&
  i[\hat{A}^{+}_{2}(t)\hat{A}_{1}(t)-\hat{A}^{+}_{1}(t)\hat{A}_{2}(t)]=
  Tr[-\hat{\sigma}_{2}\hat{D}^{+}_{2}(t)\hat{D}_{1}(t)]\label{S3}
\end{eqnarray}
where $\hat{\sigma}_{k}$ are Pauli matrices
($k=\overline{1,3\mathstrut}$), and $D$-matrices are:
\begin{equation}
D_{1}(t)=\left(
\begin{array}{cc}
  \hat{A}_{1}(t) & 0 \\
  0 & \hat{A}_{2}(t) \\
\end{array}\right),\quad
D_{2}(t)=\left(
\begin{array}{cc}
  0 & \hat{A}_{1}(t) \\
 \hat{A}_{2}(t) & 0 \\
\end{array}\right).
\end{equation}
The definition of Stokes operators $\hat{S}_{k}$ in terms of Pauli
matrices (\ref{S1})-(\ref{S3}) suggests that in quantum optics they
play a similar role to the Pauli matrices $\hat{\sigma}_{k}$ which
in quantum theory of angular momentum describe the rotations in
two-spinor formalism. Indeed, apart from a normalization factor, the
commutation relations for Stokes operators are identical to those
for Pauli matrices;
$[\hat{S}_{i}(t),\hat{S}_{j}(t)]=2i\varepsilon_{ijk}\hat{S}_{k}(t)$,
where indices $i,j,k=1,2,3$ are taken by cyclic permutations. The
energy operator $\hat{S}_{0}(t)$ commutes with any $\hat{S}_{k}(t)$,
i.e., $[\hat{S}_{0}(t),\hat{S}_{k}(t)]=0$. In addition, the Stokes
operators generate a special non-abelian unitary group of symmetry
transformations SU2 that obeys the same Lie algebra as the
three-dimensional rotation group of Pauli matrices. Since the Stokes
operator $\hat{S}_{0}(t)$ describes the total energy of the two
pulse field at the time $t$, $\hat{S}_{k}(t)$ characterize light
polarization and form a Cartesian axis system. The polarization
state is visualized as a vector on the Poincar\'e sphere and each
point on the sphere corresponds to a definite polarization state
whose variation is characterized by the motion of the point on the
sphere. If the Stokes vector points in the direction of
$\hat{S}_{1}$, $\hat{S}_{2}$, or $\hat{S}_{3}$, than the polarized
part of the beam is horizontally, linearly at $45^{\circ}$, or
right-circularly polarized, respectively. When Stokes vectors of two
USPs point in opposite directions pulses do not interfere. The
radius of Poincar\'e sphere, i.e., the average length of the
Poincar\'e vector
${\mathcal{R}}(t)=(\sum_{k=1}^{3}\langle\hat{S}_{k}(t)\rangle^{2})^{1/2}$,
defines the average intensity of the polarized part of the
radiation. The ratio of the intensity of the polarized part to the
total average intensity $\langle\hat{S}_{0}(t)\rangle$ is called
degree of polarization
${\mathcal{P}}(t)={{\mathcal{R}}(t)}/\langle\hat{S}_{0}(t)\rangle$
and it is an important measure in quantum optics.

To analyze the behavior of quantum fluctuations of Stokes operators
$\hat{S}_{\alpha}$ we introduce their correlation functions:
\begin{equation}
R_{S_\alpha}(t,t+\tau)=\langle\hat{S}_{\alpha}(t)\hat{S}_{\alpha}(t+\tau)\rangle-
\langle\hat{S}_{\alpha}(t)\rangle\langle\hat{S}_{\alpha}(t+\tau)\rangle.\label{c-function}
\end{equation}
In the frame of the quantum model for the SPM effect presented in
Section \ref{Section1} the correlation functions (\ref{c-function})
can be analytically computed by using permutation relations of the
type (\ref{permut}) and the averages (\ref{average1}) and
(\ref{average2}) of time-dependent Bose operators (see for example
\cite{EPL}). The average values in (\ref{c-function}) are calculated
on the summary coherent quantum state
$|\alpha(t)\rangle=|\alpha_{1}(t)\rangle\otimes|\alpha_{2}(t)\rangle$.
We assume here that the initial USPs are in coherent states, the
$j$-th pulse's operator $\hat{A}_{j}(t)$ acting only on the
corresponding state vector $|\alpha_{j}(t)\rangle$ within the
associated sub-Hilbert space ${{\mathcal{H}}_{j}}$, so the
factorization of the quantum states of the two sub-Hilbert spaces
takes place. The orthogonal basis of summary states belongs to the
global Hilbert space ${\mathcal{H}}=
{{\mathcal{H}}_{1}}\otimes{{\mathcal{H}}_{2}}$. This basic
assumption is followed in the paper when calculating the quantum
average of physical quantities of interest. If both USPs are in
coherent states, then $R_{S_{\alpha}}(t,t+\tau)=\delta(\tau)$. As is
well known from experiments, if USPs undergo some nonlinear
transformations (for example, parametric amplification or SPM
effect) and further overlap spatially, than the formation of PS
state of light is allowed. This means that, in the theoretical
description of the formation of the PS state in such processes, the
correlation functions of Stokes parameters are different from
$\delta$-function. However, the polarization squeezing cannot be
obtained in all Stokes parameters at the same time.  We show bellow,
by analytically deriving the correlation functions of Stokes
parameters that the spatial overlap of quadrature-squeezed pulses,
the overlap of nonclassical pulses inside an anisotropic Kerr medium
in the presence of combined SPM-XPM effect, or the interference of
nonclassical USPs on a BS, lead to the formation of the PS state.

With the correlation function calculated in agreement with
(\ref{c-function}) one can proceed with the evaluation of the
spectra $S_{S_{\alpha}}(\omega,t)$ of quantum fluctuations of Stokes
operators at any instant $t$ by merely applying the
Wiener-Khintchine (WK) theorem, which states that
\begin{equation}\label{WK}
S_{S_{\alpha}}(\omega,t)=\int^{\infty}_{-\infty}R_{S_{\alpha}}(t,t+\tau)e^{i\omega\tau}d\tau.
\end{equation}
For two coherent USPs that overlap spatially the spectrum of quantum
fluctuations of $\hat{S}_{\alpha}$ is frequency independent and
constant, i.e., $S^{coh}_{S_{\alpha}}(\omega,t)=1$. Consequently,
the deviation from the coherent level
$S^{*}_{S_{\alpha}}(\omega,t)=S_{S_{\alpha}}(\omega,t)-1$ can be
used as a measure of quantum fluctuations behavior. Thus, since the
case $0< S^{*}_{S_{\alpha}}(\omega,t)$ corresponds to
anti-squeezing, the case $-1\leq S^{*}_{S_{\alpha}}(\omega,t)<0$
corresponds to the suppression of quantum fluctuations of
$\hat{S}_{\alpha}$.
%----------------------------------------------------------------
\section{Overlapping coherent and quadrature-squeezed
pulses}\label{Section3}

\begin{figure}
\centering
\includegraphics[height=.5\textwidth]{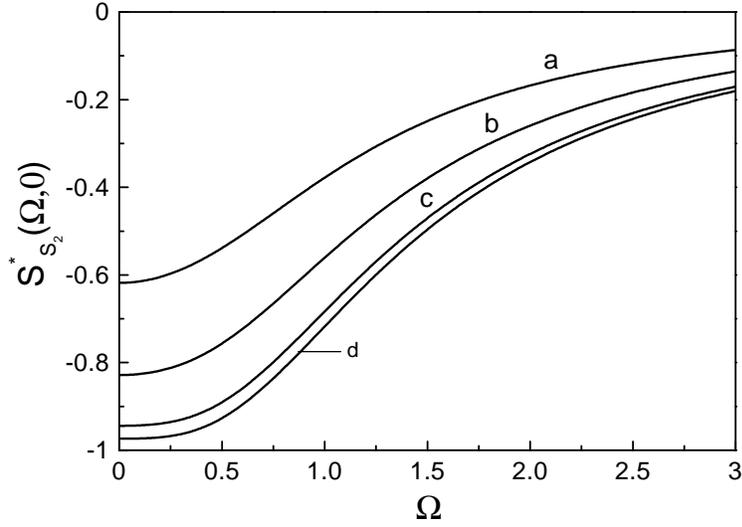}
\caption{Normalized spectral variance $S^{*}_{S_{2}}(\Omega,t)$ for
linear phase difference $\Delta\varphi(t)$ chosen optimal at
$\Omega_{0}=0$. Curves are calculated at time $t=0$ and correspond
to $\phi_{0,2}=0.5$ (a), $\phi_{0,2}=1$ (b), $\phi_{0,2}=2$ (c),
$\phi_{0,2}=3$ (d).}\label{Fig1}
\end{figure}

\begin{figure}
\centering
\includegraphics[height=.5\textwidth]{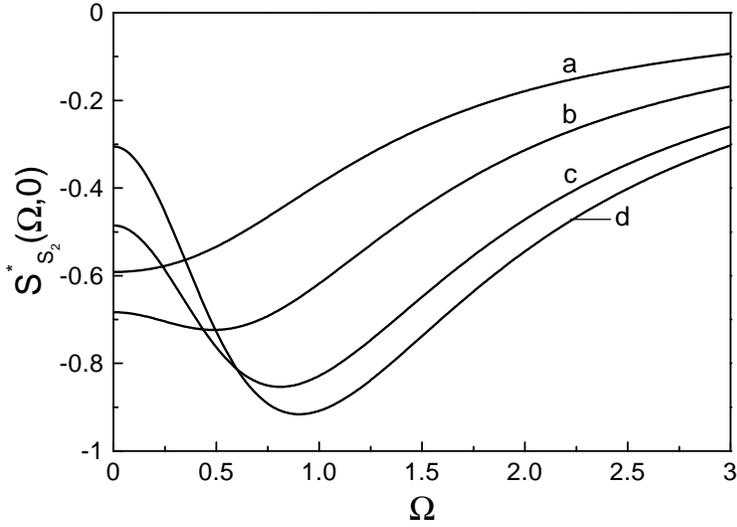}
\caption{As in Fig.\ \ref{Fig1} but for $\Omega_{0}=1$.}
\label{Fig2}
\end{figure}

In the quantum model for the SPM effect in Kerr media developed in
\cite{POP99,POP00} and shortly exposed in Section \ref{Section1} the
finite relaxation time of the electronic nonlinearity is accounted.
Note that this approach leads to an accurate calculation of
correlation functions and fluctuation spectra of quadrature
components for USPs subjected to the combined SPM-XPM effect in
electronic Kerr media \cite{QSO}. Within the quantum model
\cite{POP99,POP00} we consider the spatial overlap of a coherent USP
($\hat{A}_{1}(t)$) with a quadrature-squeezed one
($\hat{A}_{2}(t,l)$). The nonclassical USP is obtained by passing a
Kerr medium of length $l$ [$\gamma=\beta l$,
$\hat{A}_{2}(t,l)=e^{\hat{O}(t)}\hat{A}_{2}(t)$]. The Stokes
operators are defined in agreement with (\ref{S0})-(\ref{S3}) with
D-matrices in the form:
\begin{equation}
D_{1}(t)=\left(%
\begin{array}{cc}
  \hat{A}_{1}(t) & 0 \\
  0 & \hat{A}_{2}(t,l) \\
\end{array}%
\right),\quad D_{2}(t)=\left(%
\begin{array}{cc}
  0 & \hat{A}_{1}(t) \\
  \hat{A}_{2}(t,l) & 0 \\
\end{array}%
\right).
\end{equation}
We point out that the Kerr effect does not affect the photon
statistics \cite{Kitagawa} and, in the assumption that one can
neglect the dispersion effects, the operators $\hat{S}_{0}$ and
$\hat{S}_{1}$, as well as their dispersions, are conserved. Below,
by investigating the fluctuation spectra, we show that the overlap
of USPs produces the squeezing of quantum fluctuations in
$\hat{S}_{2}$ or $\hat{S}_{3}$. With the use of Eqs.\ (\ref{anih})
and (\ref{average1}) we estimate the average values of these
operators. We have: $\langle
\hat{S}_{0}(t)\rangle=\sum_{i=1}^{2}\bar{n}_{i}(t)$, $\langle
\hat{S}_{1}(t)\rangle=\bar{n}_{1}(t)-\bar{n}_{2}(t)$,
\begin{eqnarray}
\langle\hat{S}_{2}(t)\rangle &=& 2[\bar{n}_{1}(t)\bar{n}_{2}(t)]^{1/2}e^{-\mu_{2}(t)}\cos{[\Phi_{2}(t)-\varphi_{1}(t)]},\label{av1} \\
\langle\hat{S}_{3}(t)\rangle &=&
2[\bar{n}_{1}(t)\bar{n}_{2}(t)]^{1/2}e^{-\mu_{2}(t)}\sin{[\Phi_{2}(t)-\varphi_{1}(t)]},\label{av2}
\end{eqnarray}
where $\Phi_{2}(t)=\phi_{2}(t)+\varphi_{2}(t)$.  Note that
$\langle\hat{S}_{3}(t)\rangle$ is shifted in phase with $-\pi/2$ in
comparison with $\langle\hat{S}_{2}(t)\rangle$, in complete
agreement with the Heisenberg uncertainty relation for noncommuting
observables. To investigate the behavior of quantum fluctuation of
Stokes operators $\hat{S}_{2}$  and $\hat{S}_{2}$  we proceed by
evaluating their correlation functions (\ref{c-function}). It is not
difficult to show that in the approximation $\gamma\ll 1$ by using
Eqs.\ (\ref{anih})-(\ref{average2}) the correlation function
$R_{S_{2}}(t,t+\tau)$ has the following analytical form:
\begin{eqnarray}
R_{S_{2}}(t,t+\tau)=\delta(\tau)&+&h(\tau)\bar{n}_{1}(t)\phi_{2}(t)\sin{2[\varphi_{1}(t)-\Phi_{2}(t)]}\nonumber\\
&+&g(\tau)\bar{n}_{1}(t)\phi^{2}_{2}(t)\sin^{2}{[\varphi_{1}(t)-\Phi_{2}(t)]}.\label{c1}
\end{eqnarray}
With the correlation function (\ref{c1}) the fluctuation spectrum of
$\hat{S}_{2}$ can be obtained by applying the WK theorem (\ref{WK}).
Allowing for a small change of the envelope $r(t)$ during the
relaxation time $\tau_{r}$ and accounting the fact that for the
electronic nonlinearity (\ref{relaxation}) we have:
$\int_{-\infty}^{\infty}h(\tau)e^{i\omega\tau}d\tau=2L(\omega)$ and
$\int_{-\infty}^{\infty}g(\tau)e^{i\omega\tau}d\tau=4L^{2}(\omega)$
\cite{POP99,POP00,QSO}, for the fluctuation spectrum of
$\hat{S}_{2}$ we finally obtain
\begin{eqnarray}
S_{S_{2}}(\Omega,t)=1&+&2L(\Omega)\bar{n}_{1}(t)\phi_{2}(t)\sin{2[\varphi_{1}(t)-\Phi_{2}(t)]}\nonumber\\
&+&4L^{2}(\Omega)\bar{n}_{1}(t)\phi^{2}_{2}(t)\sin^{2}{[\varphi_{1}(t)-\Phi_{2}(t)]},\label{sp1}
\end{eqnarray}
where $L(\Omega)=1/(1+\Omega^{2})$. Here the dimensionless frequency
$\Omega=\omega\tau_{r}$ is introduced for the simplicity of
notations. The correlation function of $\hat{S}_{3}$, as well as its
fluctuation spectrum, can be easily obtained by shifting (\ref{c1})
and (\ref{sp1}) in phase with $\pi/2$, respectively. Since the
spectral density of quantum fluctuations (\ref{sp1}) depends on the
relaxation time of the electronic nonlinearity, the latter defines
the level of quantum fluctuations below to the one corresponding to
the coherent state. The second term on the r.h.s. of Eq.\
(\ref{sp1}), which can be negative for a definite linear phase
difference $\Delta\varphi(t)=\varphi_{2}(t)-\varphi_{1}(t)$ between
USPs, indicates clearly that the level of quantum fluctuations of
$\hat{S}_{2}$ can be less than that corresponding to the coherent
state, $S_{S_{2}}^{coh}(\Omega,t)=1$. Let $\Delta\varphi(t)$ at a
definite reduced frequency $\Omega_{0}$ has the form
\begin{equation}\label{opt-QS-CH}
\Delta\varphi_{opt}(t)=\frac{1}{2}\arctan{\left(\frac{1}{L(\Omega_{0})\phi_{2}(t)}\right)}-\phi_{2}(t).
\end{equation}
Then, at this $\Delta\varphi_{opt}(t)$ the fluctuation spectrum
(\ref{sp1}) reaches the minimum value
\begin{equation}\label{spectrumA-QS-CH}
S_{S_{2}}(\Omega_{0},t)=1+2\bar{n}_{1}(t)\phi^{2}_{2}(t)L^{2}(\Omega_{0})-2\bar{n}_{1}(t)\phi_{2}(t)L(\Omega_{0})[1+\phi^{2}_{2}(t)L^{2}(\Omega_{0})]^{1/2}.
\end{equation}
At any frequency $\Omega$ the fluctuation spectrum writes:
\begin{eqnarray}\label{spectrumB-QS-CH}
S_{S_{2}}(\Omega,t)&=&S_{S_{2}}(\Omega_{0},t)+2\bar{n}_{1}(t)\phi^{2}_{2}(t)[L^{2}(\Omega)-L^{2}(\Omega_{0})]-2\bar{n}_{1}(t)
\phi_{2}(t)[L(\Omega)-L(\Omega_{0})]\nonumber\\
&{}&\times\{1+\phi^{2}_{2}(t)L(\Omega_{0})[L(\Omega)+L(\Omega_{0})]\}[1+\phi_{2}^{2}(t)L^{2}(\Omega)]^{-1/2}
\end{eqnarray}
As already pointed out, in order to characterize the deviation from
the coherent level we introduce the normalized spectral variance
$S^{*}_{S_{2}}(\Omega,t)=[S_{S_{2}}(\Omega,t)-1]/\bar{n}_{1}(t)$. In
Figs.\ \ref{Fig1} and \ref{Fig2} we display
$S^{*}_{S_{2}}(\Omega,t)$ for different values of the nonlinear
phase addition $\phi_{2}(t)$. For linear phase difference optimized
at low frequencies $\Omega_{0}=0$ the squeezing in $\hat{S}_{2}$ is
maximal around $\Omega\approx0$ (Fig.\ \ref{Fig1}). In case
$\Omega_{0}=1$, the squeezing take place essentially at high
frequencies $\Omega\approx1$ ($\omega=1/\tau_{r}$) (Fig.\
\ref{Fig2}). Thus, the choice of linear phase difference allow us to
produce the squeezing basically at the frequency of interest.
Moreover, the increase of the nonlinear phase addition $\phi_{2}(t)$
produces a better level of squeezing.
%%%%%%%%%%%%%%%%%%%%%%%%%%%%%%%%%%%%%%%%%%%%%%%%%%%%%%%%%%%%%%%%%%%%%%%%%%%
%%%%%%%%%%%%%%%%%%%%%%%%%%%%%%%%%%%%%%%%%%%%%%%%%%%%%%%%%%%%%%%%%%%%%%%%%%
%%%%%%%%%%%%%%%%%%%%%%%%%%%%%%%%%%%%%%%%%%%%%%%%%%%%%%%%%%%%%%%%%%%%%%%%%%%
%%%%%%%%%%%%%%%%%%%%%%%%%%%%%%%%%%%%%%%%%%%%%%%%%%%%%%%%%%%%%%%%%%%%%%%%%%
%----------------------------------------------------------------
\section{Overlapping quadrature-squeezed pulses}\label{Section4}
\begin{figure}
\centering
\includegraphics[height=.5\textwidth]{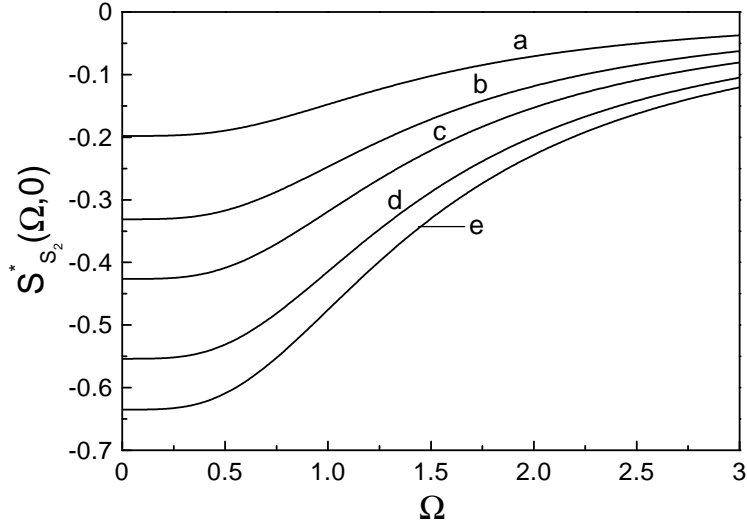}
\caption{Normalized spectral variance $S^{*}_{S_{2}}(\Omega,t)$ for
linear phase difference $\Delta\varphi(t)$ chosen optimal at
$\Omega_{0}=0$. Curves are calculated at time $t=0$ with
$\phi_{1,0}=2$ and $\gamma_{1}=2\gamma_{2}$ and correspond to
$\bar{n}_{2,0}/\bar{n}_{1,0}=1$ (a), $\bar{n}_{2,0}/\bar{n}_{1,0}=2$
(b), $\bar{n}_{2,0}/\bar{n}_{1,0}=3$ (c),
$\bar{n}_{2,0}/\bar{n}_{1,0}=5$ (d), $\bar{n}_{2,0}/\bar{n}_{1,0}=7$
(e).} \label{Fig3}
\end{figure}

\begin{figure}
\centering
\includegraphics[height=.5\textwidth]{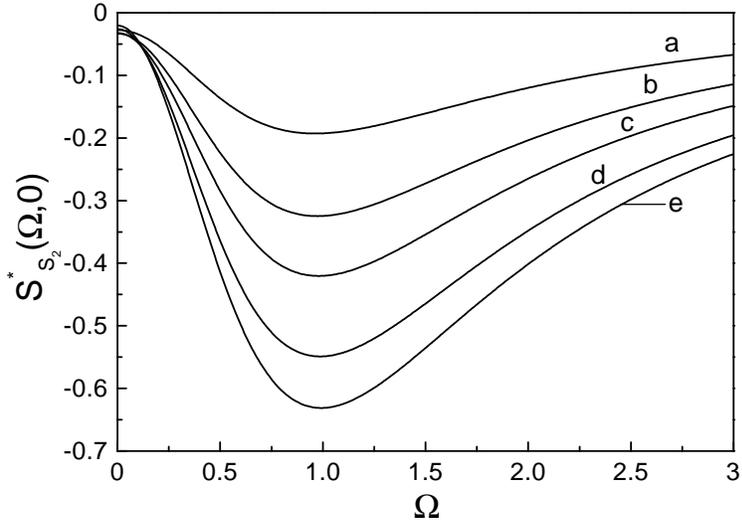}
\caption{As in Fig.\ \ref{Fig3} but for $\Omega_{0}=1$.}
\label{Fig4}
\end{figure}

Now we focus our attention to the case of spatial overlap of two
independent quadrature-squeezed USPs. The nonclassical pulses
obtained at the exits of electronic Kerr media of lengths $l_{1}$
and $l_{2}$ further overlap spatially. The Stokes operators for the
nonclassical USPs are defined in agreement with
(\ref{S0})-(\ref{S3}) in terms of annihilation
[$\hat{A}_{j}(t,l_{j})=e^{\hat{O}_{j}(t)}\hat{A}_{j}(t)$] and
creation
[$\hat{A}^{+}_{j}(t,l_{j})=\hat{A}^{+}_{j}(t)e^{\hat{O}^{+}_{j}(t)}$]
operators ($j=1,2$) with D-matrices in the form:
\begin{equation}
D_{1}(t)=\left(%
\begin{array}{cc}
  \hat{A}_{1}(t,l_{1}) & 0 \\
  0 & \hat{A}_{2}(t,l_{2}) \\
\end{array}%
\right),\quad D_{2}(t)=\left(%
\begin{array}{cc}
  0 & \hat{A}_{1}(t,l_{1}) \\
  \hat{A}_{2}(t,l_{2}) & 0 \\
\end{array}%
\right).
\end{equation}
Proceeding as in the previous section for the average value of
$\hat{S}_{2}$ and $\hat{S}_{3}$ we have:
\begin{eqnarray}
\langle\hat{S}_{2}(t)\rangle &=& 2[\bar{n}_{1}(t)\bar{n}_{2}(t)]^{1/2}
e^{-\mu_{1}(t)-\mu_{2}(t)}\cos{[\Phi_{2}(t)-\Phi_{1}(t)]},\label{avv1} \\
\langle\hat{S}_{3}(t)\rangle &=&
2[\bar{n}_{1}(t)\bar{n}_{2}(t)]^{1/2}
e^{-\mu_{1}(t)-\mu_{2}(t)}\sin{[\Phi_{2}(t)-\Phi_{1}(t)]}.\label{avv2}
\end{eqnarray}
The correlation function of $\hat{S}_{2}$ calculated by using the
relations (\ref{anih})-(\ref{average2}) in the approximation
$\gamma_{j}\ll 1$ is
\begin{eqnarray}
R_{S_{2}}(t,t+\tau)=\delta(\tau)&+&h(\tau)[\bar{n}_{1}(t)\phi_{2}(t)-\bar{n}_{2}(t)\phi_{1}(t)]\sin{2[\Phi_{1}(t)-\Phi_{2}(t)]}\nonumber\\
&+&g(\tau)[\bar{n}_{1}(t)\phi^{2}_{2}(t)+\bar{n}_{2}(t)\phi^{2}_{1}(t)]\sin^{2}{[\Phi_{1}(t)-\Phi_{2}(t)]}.\label{c2}
\end{eqnarray}
Compare the relations (\ref{av1})-(\ref{c1}) and
(\ref{avv1})-(\ref{c2}). Instead of $\varphi_{1}(t)$ we have now
$\Phi_{1}(t)=\phi_{1}(t)+\varphi_{1}(t)$ where $\phi_{1}(t)$ is the
nonlinear phase addition due to SPM effect on the pulse $\textbf{1}$
after passing the Kerr medium of length $l_{1}$. With the WK
theorem, for the fluctuation spectrum of $\hat{S}_{2}$ we get:
\begin{eqnarray}
S_{S_{2}}(\Omega,t)=1&+&2L(\Omega)[\bar{n}_{1}(t)\phi_{2}(t)-\bar{n}_{2}(t)\phi_{1}(t)]\sin{2[\Phi_{1}(t)-\Phi_{2}(t)]}\\
&+&4L^{2}(\Omega)[\bar{n}_{1}(t)\phi^{2}_{2}(t)+\bar{n}_{2}(t)\phi^{2}_{1}(t)]\sin^{2}{[\Phi_{1}(t)-\Phi_{2}(t)]}.\label{spp1}
\end{eqnarray}
Besides adjusting the nonlinear phase differences $\phi_{j}(t)$, we
can control the spectral density (\ref{spp1}) by changing the
intensities $\bar{n}_{j}(t)$ of USPs. If the linear phase difference
$\Delta\varphi(t)=\varphi_{2}(t)-\varphi_{1}(t)$ at a definite
frequency $\Omega_{0}$ is taken in the form
\begin{equation}\label{opt-2QS}
\Delta\varphi_{opt}=\frac{1}{2}\arctan{\left(\frac{\bar{n}_{1}(t)\phi_{2}(t)-
\bar{n}_{2}(t)\phi_{1}(t)}{L(\Omega_{0})[\bar{n}_{1}(t)\phi^2_{2}(t)+\bar{n}_{2}(t)\phi^2_{1}(t)]}\right)}+\phi_{1}(t)-\phi_{2}(t),
\end{equation}
then the fluctuation spectrum (\ref{spp1}) reaches the minimum value
\begin{eqnarray}\label{spectrumA-2QS}
S_{S_{2}}(\Omega_{0},t)&=&1+2[\bar{n}_{1}(t)\phi^{2}_{2}(t)+
\bar{n}_{2}(t)\phi^2_{1}(t)]L^{2}(\Omega_{0})-
2L(\Omega_{0})\{[\bar{n}_{1}(t)\phi_{2}(t)-\bar{n}_{2}(t)\phi_{1}(t)]^2\nonumber\\
&+&L^{2}(\Omega_0)
[\bar{n}_{1}(t)\phi^2_{2}(t)+\bar{n}_{2}(t)\phi^2_{1}(t)]^2\}^{1/2}.
\end{eqnarray}
At any frequency $\Omega$ the fluctuation spectrum is:
\begin{eqnarray}\label{spectrumB-2QS}
S_{S_{2}}(\Omega,t)\!&=&\!S_{S_{2}}(\Omega_{0},t)+2[\bar{n}_{1}(t)\phi^{2}_{2}(t)+
\bar{n}_{2}(t)\phi^{2}_{1}(t)][L^{2}(\Omega)-L^{2}(\Omega_{0})]-2[L(\Omega)-L(\Omega_{0})]\nonumber\\
&\times&\!\!\{\left[\bar{n}_{1}(t)\phi_{2}(t)-\bar{n}_{2}(t)\phi_{1}(t)\right]^2
+L(\Omega_{0})[L(\Omega)+L(\Omega_{0})][\bar{n}_{1}(t)\phi^2_{2}(t)+
\bar{n}_{2}(t)\phi^2_{1}(t)]^2\}\nonumber\\
&\times&\!\!\left\{[\bar{n}_{1}(t)\phi_{2}(t)-\bar{n}_{2}(t)\phi_{1}(t)]^2+L^{2}(\Omega_0)
[\bar{n}_{1}(t)\phi^2_{2}(t)+\bar{n}_{2}(t)\phi^2_{1}(t)]^2\right\}^{-1/2}.
\end{eqnarray}
In Figs.\ \ref{Fig3} and \ref{Fig4} we present the normalized
spectral variance
$S^{*}_{S_{2}}(\Omega,t)=[S^{coh}_{S_{2}}(\Omega,t)-1]/\bar{n}_{1}(t)$
for different relations between pulses intensities. One can see
that, by increasing  the intensity of the control USP $\mathbf{2}$,
we achieve a better level of squeezing.
%%%%%%%%%%%%%%%%%%%%%%%%%%%%%%%%%%%%%%%%%%%%%%%%%%%%%%%%%%%%%%%%%%%%%%%%%%%
%%%%%%%%%%%%%%%%%%%%%%%%%%%%%%%%%%%%%%%%%%%%%%%%%%%%%%%%%%%%%%%%%%%%%%%%%%
%%%%%%%%%%%%%%%%%%%%%%%%%%%%%%%%%%%%%%%%%%%%%%%%%%%%%%%%%%%%%%%%%%%%%%%%%%%
%%%%%%%%%%%%%%%%%%%%%%%%%%%%%%%%%%%%%%%%%%%%%%%%%%%%%%%%%%%%%%%%%%%%%%%%%%
\section{Overlapping quadrature-squeezed USPs in anisotropic Kerr media}\label{Section5}
When two USPs with orthogonal polarization and/or different
frequencies overlap spatially inside an anisotropic Kerr medium,
beside SPM effect of pulses, the XPM effect and parametric
interaction occur. The parametric frequency conversion can be
neglected in the assumption of a large phase mismatch
$|\Delta|>\rm{min}\{\tilde{\beta}|A_{1}(0)|,\tilde{\beta}|A_{2}(0)|\}$
and $\Delta z\gg1$ \cite{Orlov,Ahmanov}. In this situation the XPM
effect can be employed to control the spectra of quantum
fluctuations of quadratures \cite{QSO}, as well as those of quantum
Stokes parameters \cite{EPL}. The consistent description of the
combined SPM-XPM effect is based on the momentum operator
$\hat{G}(z)=\sum_{j=1}^{2}\hat{G}_{\rm{spm}}(z)+\hat{G}_{\rm{xpm}}(z)$
where
\begin{equation}
\hat{G}_{\rm{xpm}}(z)=\hbar\widetilde{\beta}\int_{-\infty}^{\infty}dt
\int_{-\infty}^{t}H(t-t_{1})\left[\hat{n}_{1}(t,z)\hat{n}_{2}(t_{1},z)+
\hat{n}_{1}(t_{1},z)\hat{n}_{2}(t,z)\right]dt_{1}. \label{xpm}
\end{equation}
In the above expression $\tilde{\gamma}=\widetilde{\beta}z$ is the
nonlinear coupling coefficient \cite{QSO,EPL}. With the total
momentum operator $\hat{G}(z)$ we derive the solution of the
space-evolution equation (\ref{part1}) for the pulse $\mathbf{1}$;
\begin{equation}\label{anihXPM}
\hat{A}_{1}(t,z)=e^{\hat{O}_{1}(t)+\hat{\tilde{O}}_{2}(t)}\,\hat{A}_{1}(t),
\end{equation}
where $\hat{\tilde{O}}_{j}(t)=i\tilde{\gamma}q(\hat{n}_{j}(t))$
includes the contribution due to the XPM effect. The expression for
$\hat{A}_{2}(t,z)$ can be derived by changing the indexes
$1\leftrightarrow 2$ in (\ref{anihXPM}). The application of the
solution (\ref{anihXPM}) requires the use of the extended algebra of
time-dependent Bose operators that contains permutation relations of
the form
\begin{eqnarray}
\hat{A}_{j}(t_{1})e^{\hat{O}_{j}(t_{2})}&=&e^{\hat{O}_{j}(t_{2})
+{\mathcal{D}}_{j}(t_{2}-t_{1})}\hat{A}_{j}(t_{1}),\label{permutXPM1}\\
\hat{A}_{j}(t_{1})e^{\hat{\tilde{O}}_{j}(t_{2})}&=&e^{\hat{\tilde{O}}_{j}(t_{2})
+{\tilde{\mathcal{D}}}(t_{2}-t_{1})}\hat{A}_{j}(t_{1}),\label{permutXPM2}
\end{eqnarray}
where here
${\tilde{\mathcal{D}}}(t_{2}-t_{1})=i\tilde{\gamma}h(t_{2}-t_{1})$
\cite{QSO}. With expressions (\ref{permutXPM1})-(\ref{permutXPM2})
for the average values of the Bose operators over the initial
coherent summary state $|\alpha(t)\rangle$ we get:
\begin{eqnarray}
\langle e^{\hat{O}_{j}(t)}\rangle &=&
e^{i\phi_{j}(t)-\mu_{j}(t)},\quad ~\langle
e^{\hat{\tilde{O}}_{j}(t)}\rangle
=e^{i\tilde{\phi}_{j}(t)-\tilde{\mu}_{j}(t)},\label{hat1}\\
\langle e^{\hat{O}_{j}(t_{1})+\hat{O}_{j}(t_{2})}\rangle &=&
e^{i[\phi_{j}(t_{1})+\phi_{j}(t_{2})]-[\mu_{j}(t_{1})+\mu_{j}(t_{2})]-
{\mathcal{K}}_{j}(t_{1},t_{2})},\label{susi1}\\ \langle
e^{\hat{\tilde{O}}_{j}(t_{1})+\hat{\tilde{O}}_{j}(t_{2})}\rangle &=&
e^{i[\tilde{\phi}_{j}(t_{1})+\tilde{\phi}_{j}(t_{2})]-
[\tilde{\mu}_{j}(t_{1})+\tilde{\mu}_{j}(t_{2})]-\tilde{\mathcal{K}}_{j}(t_{1},t_{2})}.\label{tutu1}
\end{eqnarray}
The parameters $\tilde{\phi}_{j}(t)=2\tilde{\gamma}\bar{n}_{j}(t)$,
$\tilde{\mu}_{j}(t)=\tilde{\gamma}^{2}\bar{n}_{j}(t)/2$ are
connected with the XPM of the $j$-th pulse. Here
$\tilde{\phi}_{j}(t)$ and $\tilde{\mathcal{K}}_{j}(t_{1},t_{2})$ are
the nonlinear phase addition and the temporal correlator caused by
XPM, respectively.

\begin{figure}
\centering
\includegraphics[height=.5\textwidth]{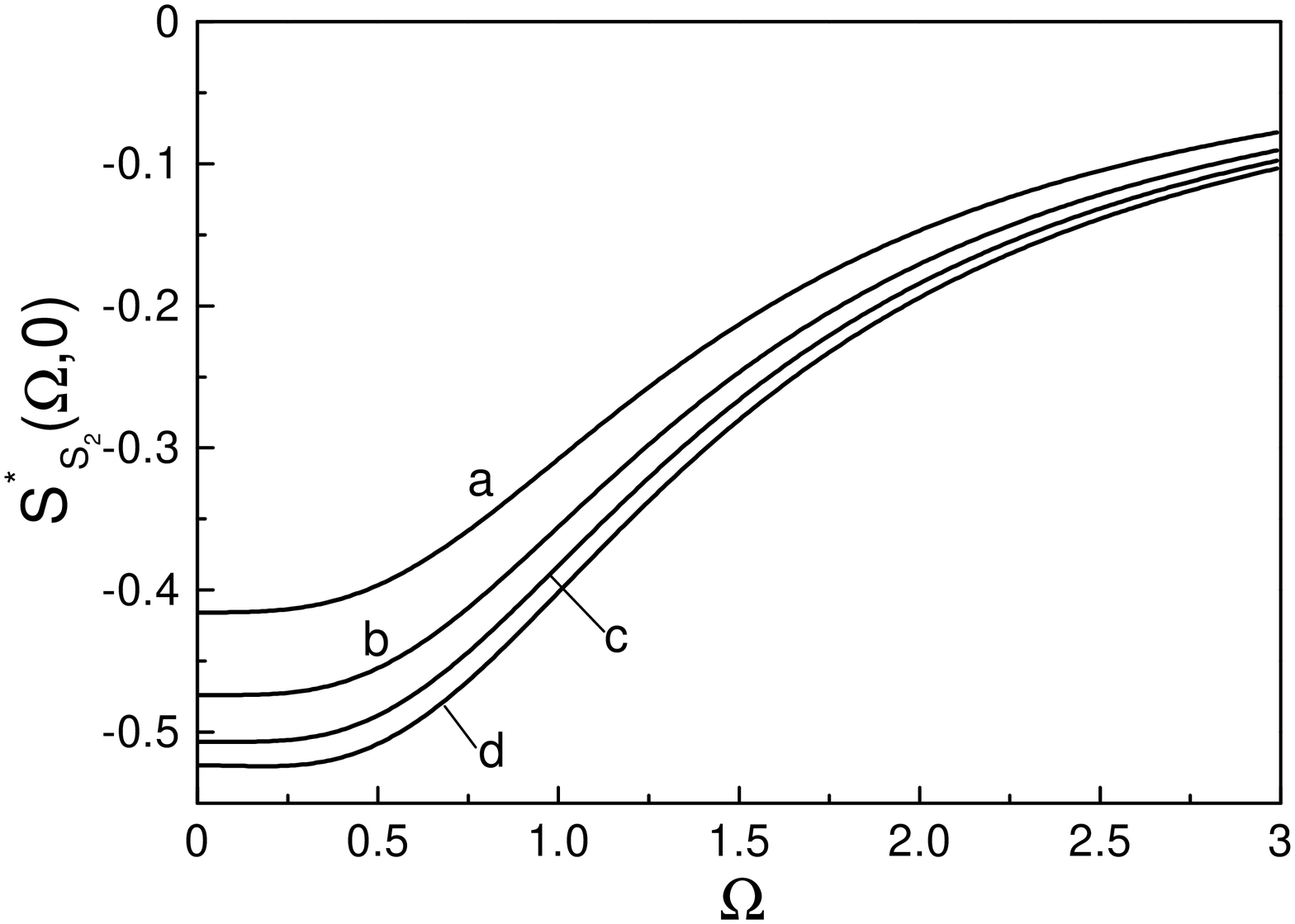}
\caption{Normalized spectral variance $S^{*}_{S_{2}}(\Omega,t)$ at
$\phi_{0,1}=2$ for initial phase difference $\Delta\varphi(t)$
chosen optimal at $\Omega_{0}=0$. Curves are calculated at time
$t=0$, $\gamma_{1}=\gamma_{2}/4=2\tilde{\gamma}$, and correspond to
$\bar{n}_{0,2}=\bar{n}_{0,1}/4$ (a), $\bar{n}_{0,2}=\bar{n}_{0,1}/2$
(b), $\bar{n}_{0,2}=\bar{n}_{0,1}$ (c),
$\bar{n}_{0,2}=3\bar{n}_{0,1}$ (d) (from Ref. [26]).}\label{Fig5}
\end{figure}

\begin{figure}
\centering
\includegraphics[height=.5\textwidth]{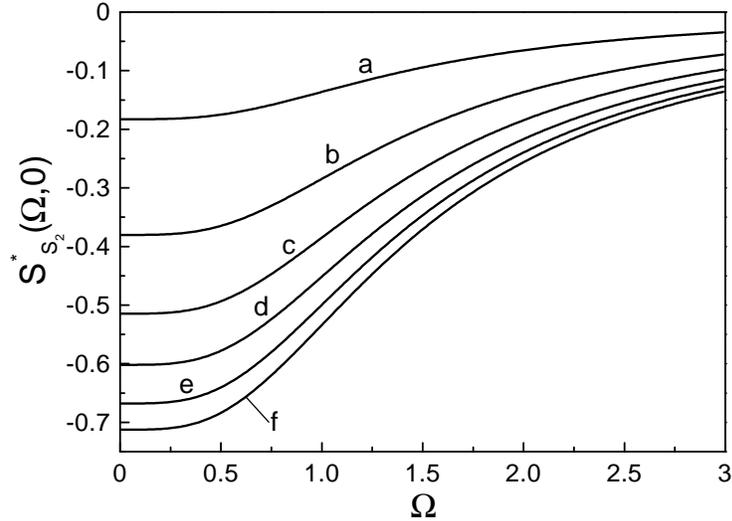}
\caption{Normalized spectral variance $S^{*}_{S_{2}}(\Omega,t)$ at
$\phi_{0,1}=2$ for initial phase difference $\Delta\varphi(t)$
chosen optimal at $\Omega_{0}=0$. Curves are calculated at time
$t=0$, $\bar{n}_{0,1}=\bar{n}_{0,2}$, $\tilde{\gamma}=\gamma_{1}/2$
and correspond to $\gamma_{2}=2\gamma_{1}$ (a),
$\gamma_{2}=3\gamma_1$ (b), $\gamma_{2}=4\gamma_1$ (c),
$\gamma_{2}=5\gamma_1$ (d), $\gamma_{2}=6\gamma_1$ (e),
$\gamma_{2}=7\gamma_1$ (f) (from Ref. [26]).} \label{Fig6}
\end{figure}

The Stokes operators are defined in agree with the formulas
(\ref{S0})-(\ref{S3}) with D-matrices in the form:
\begin{equation}
D_{1}(t)=\left(%
\begin{array}{cc}
  \hat{A}_{1}(t,z) & 0 \\
  0 & \hat{A}_{2}(t,z) \\
\end{array}%
\right),\quad D_{2}(t)=\left(%
\begin{array}{cc}
  0 & \hat{A}_{1}(t,z) \\
  \hat{A}_{2}(t,z) & 0 \\
\end{array}%
\right),
\end{equation}
where $z$ is the distance inside the anisotropic Kerr medium where
the pulses overlap. By using Eqs.\ (\ref{permutXPM1}),
(\ref{permutXPM2}) and Eq.\ (\ref{hat1}) we estimate the average
values of Stokes operators:
\begin{eqnarray}
\langle\hat{S}_{2}(t)\rangle &=&
2{[\bar{n}_{1}(t)\bar{n}_{2}(t)]}^{1/2}e^{-\Delta_{1}(t)-\Delta_{2}(t)}
\cos{[\widetilde{\Phi}_{2}(t)-\widetilde{\Phi}_{1}(t)]},\label{med2}\\
\langle\hat{S}_{3}(t)\rangle &=&
2{[\bar{n}_{1}(t)\bar{n}_{2}(t)]}^{1/2}e^{-\Delta_{1}(t)-\Delta_{2}(t)}
\sin{[\widetilde{\Phi}_{2}(t)-\widetilde{\Phi}_{1}(t)]},\label{med3}
\end{eqnarray}
where here $\Delta_{j}(t)=\mu_{j}(t)+\tilde{\mu}_{j}(t)$,
$\tilde{\Phi}_{j}(t)=\phi_{j}(t)-\tilde{\phi}_{j}(t)+\varphi_{j}(t)$.
In the approximation $\gamma_{j},\tilde{\gamma}\ll 1$, by making use
of Eqs.\ (\ref{permutXPM1}), (\ref{permutXPM2}) and Eqs.\
(\ref{susi1}), (\ref{tutu1}), for the correlation function of
$\hat{S}_{2}$ we get
\begin{eqnarray}
R_{S_{2}}(t,t+\tau) &=& \delta(\tau)+
h(\tau)[\bar{n}_{1}(t)\phi_{2}(t)-\bar{n}_{2}(t)\phi_{1}(t)]
\sin{2[\widetilde{\Phi}_{1}(t)-\widetilde{\Phi}_{2}(t)]}\nonumber\\
&+&g(\tau)\{\bar{n}_{1}(t)[\phi^{2}_{2}(t)\!+\!\tilde{\phi}^{2}_{2}(t)]+
\bar{n}_{2}(t)[\phi^{2}_{1}(t)\!+\!\tilde{\phi}^{2}_{1}(t)]\}
\sin^{2}{[\widetilde{\Phi}_{1}(t)-\widetilde{\Phi}_{2}(t)]}.\label{re1}
\end{eqnarray}
One can easily see that in the absence of XPM, i.e.
$\tilde{\phi}_{1}(t)=\tilde{\phi}_{2}(t)=0$, the expression above
reduces to the expression (\ref{c2}) of the previous section. To
derive the analytical expression of the fluctuation spectrum of
$\hat{S}_{2}$ we apply the WK theorem. Thus,
\begin{eqnarray}
S_{S_{2}}(\Omega,t)&=&1+2L(\Omega)[\bar{n}_{1}(t)\phi_{2}(t)-\bar{n}_{2}(t)\phi_{1}(t)]
\sin{2[\widetilde{\Phi}_{1}(t)-\widetilde{\Phi}_{2}(t)]}\nonumber\\
&+&4L^{2}(\Omega)\{\bar{n}_{1}(t)[\phi^{2}_{2}(t)\!+\!\tilde{\phi}^{2}_{2}(t)]+
\bar{n}_{2}(t)[\phi^{2}_{1}(t)\!+\!\tilde{\phi}^{2}_{1}(t)]\}
\sin^{2}{[\widetilde{\Phi}_{1}(t)-\widetilde{\Phi}_{2}(t)]}.\label{S2Spectra}
\end{eqnarray}
Notice that the XPM effect introduces specific contributions in the
fluctuation spectrum [compare (\ref{S2Spectra}) with (\ref{spp1})].
Their change allows for an additional control of fluctuation
spectrum (see \cite{EPL}). At the linear phase difference
$\Delta\varphi(t)=\varphi_{1}(t)-\varphi_{2}(t)$
\begin{eqnarray}\label{phasedif}
\Delta\varphi(t)_{\rm opt}
&=&\frac{1}{2}\arctan\left(\frac{\bar{n}_{1}(t)\phi_{2}(t)-\bar{n}_{2}(t)\phi_{1}(t)}
{L(\Omega_{0})\Bigl\{\bar{n}_{0,1}(t)[\phi^{2}_{2}(t)+\tilde{\phi}^{2}_{2}(t)]+
\bar{n}_{2}(t)[\phi^{2}_{1}(t)+\tilde{\phi}^{2}_{1}(t)]\Bigl\}}\right)\nonumber\\
&{}&+\phi_{1}(t)-\phi_{2}(t)-\tilde{\phi}_{1}(t)+\tilde{\phi}_{2}(t),
\end{eqnarray}
optimal at a definite frequency $\Omega_{0}=\omega_{0}\tau_{r}$, the
spectrum (\ref{S2Spectra}) reaches the minimum
\begin{eqnarray}\label{S2min}
S_{S_{2}}(\Omega_{0},t)&=&1+2L^{2}(\Omega_{0})\left(\bar{n}_{1}(t)[\phi^{2}_{2}(t)+\tilde{\phi}^{2}_{2}(t)]+
\bar{n}_{2}(t)[\phi^{2}_{1}(t)+\tilde{\phi}^{2}_{1}(t)]\right)\nonumber\\
&{}&-2L(\Omega_{0})\Bigl[\bigl[\bar{n}_{1}(t)\phi_{2}(t)-\bar{n}_{2}(t)\phi_{1}(t)\bigl]^{2}\nonumber\\
&{}&+L^{2}(\Omega_{0})\Bigl(\bar{n}_{1}(t)[\phi^{2}_{2}(t)+\tilde{\phi}^{2}_{2}(t)]+
\bar{n}_{0,2}(t)[\phi^{2}_{1}(t)+\tilde{\phi}^{2}_{1}(t)]\Bigl)^{2}\Bigl]^{1/2}.
\end{eqnarray}
In Fig. \ref{Fig5} we display the fluctuation spectra of
$\hat{S}_{2}$ for different relations between the intensities of
pulses. With the increase of intensity of the pulse \textbf{2} (the
control pulse) a better level of suppression of quantum fluctuations
can be achieved. A similar result, now displayed in Fig.\
\ref{Fig6}, can be obtained by increasing the nonlinear coefficient
$\gamma_{2}$ in comparison with $\gamma_1$. In conclusion, besides
the optimal linear phase arrangement (\ref{phasedif}) that produce
the suppression of quantum fluctuation at the frequency of interest,
the increase of the intensity of the control pulse (Fig.
\ref{Fig5}), or of the nonlinear coefficient $\gamma_2$ (Fig.
\ref{Fig6}), can be employed for the achievement of desired level of
squeezing.
%%%%%%%%%%%%%%%%%%%%%%%%%%%%%%%%%%%%%%%%%%%%%%%%%%%%%%%%%%%%%%%%%%%%%%%%%%%
%%%%%%%%%%%%%%%%%%%%%%%%%%%%%%%%%%%%%%%%%%%%%%%%%%%%%%%%%%%%%%%%%%%%%%%%%%
%%%%%%%%%%%%%%%%%%%%%%%%%%%%%%%%%%%%%%%%%%%%%%%%%%%%%%%%%%%%%%%%%%%%%%%%%%%
%%%%%%%%%%%%%%%%%%%%%%%%%%%%%%%%%%%%%%%%%%%%%%%%%%%%%%%%%%%%%%%%%%%%%%
%----------------------------------------------------------
\section{Interfering quadrature-squeezed pulses}\label{Section6}
\begin{figure}
\centering
\includegraphics[height=.5\textwidth]{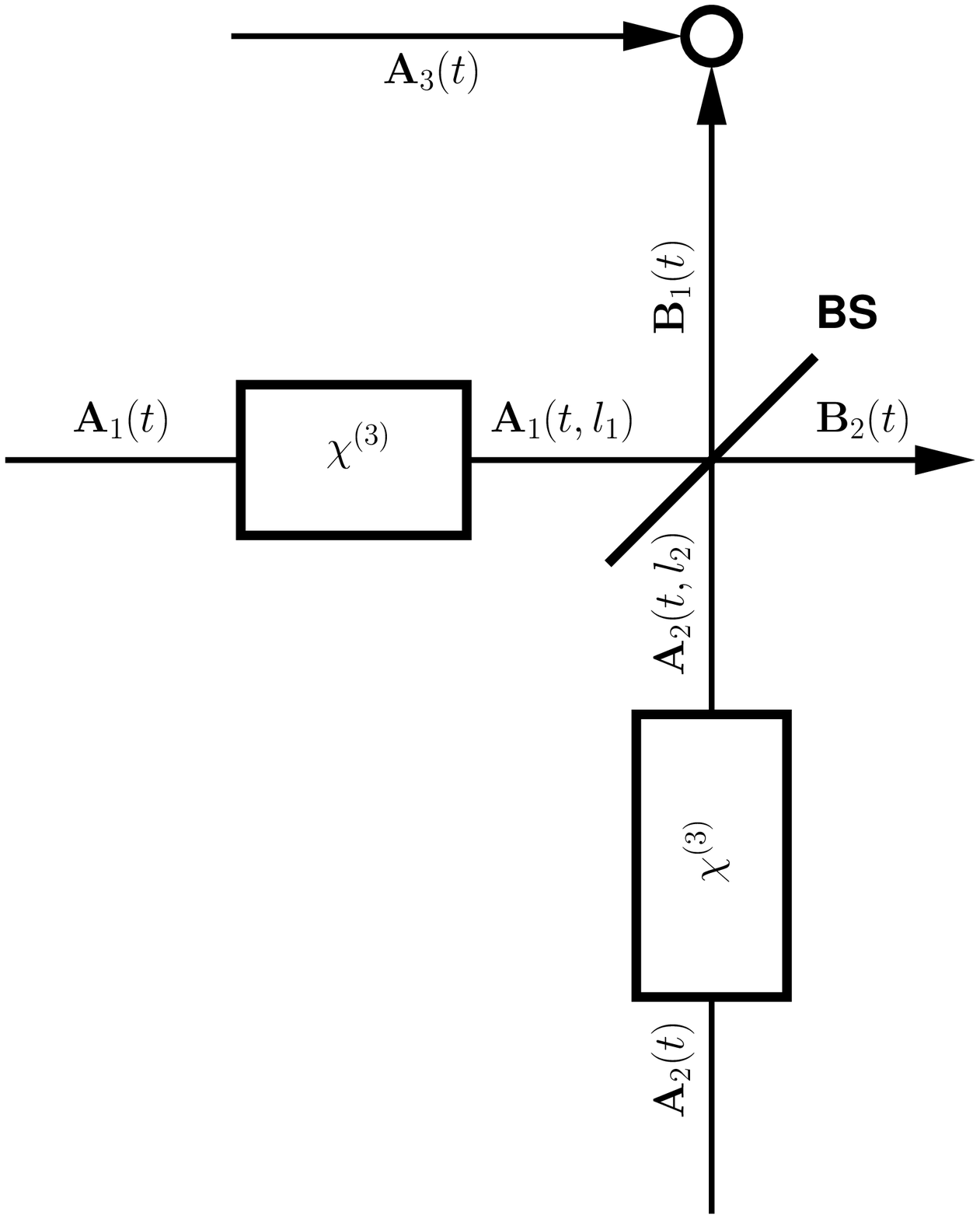}
\caption{Interference on a BS of two independent quadrature-squeezed
pulses produced by using the SPM effect in nonlinear electronic Kerr
media ($\chi^{(3)}$). The outgoing nonclassical pulse at the output
\textbf{1} ($\hat{B}_{1}(t)$) is overlapped spatially with a
coherent pulse field ($\hat{A}_{3}(t)$).}\label{Fig7}
\end{figure}

In this section we show that by overlapping a coherent pulse with an
interference USP, obtained by mixing two quadrature-squeezed USPs on
a BS, one can produce the simultaneous squeezing in the first two
Stokes parameters $\hat{S}_{0}$ and $\hat{S}_{1}$, as well as the
squeezing in one of the last two, $\hat{S}_{2}$ or $\hat{S}_{3}$. We
shall compare our theoretical results with the experimental ones
presented in \cite{Heersnik,Heersnik2}. We start by analyzing the
situation depicted in Fig.\ \ref{Fig7}.
%%%%%%%%%%%%%%%%%%%%%%%%%%%%%%%%%%%%%%%%%%%%%%%%%%%%%%%%%%%%%%%%%%%%%%%%%%%
%%%%%%%%%%%%%%%%%%%%%%%%%%%%%%%%%%%%%%%%%%%%%%%%%%%%%%%%%%%%%%%%%%%%%%%%%%
Two quadrature-squeezed USPs fall simultaneously on an inclined BS
with reflectivity $R$ and transmissivity $T$, giving rise to
reflected and transmitted interference USPs whose amplitudes depends
on $R$ and $T$. The outgoing nonclassical interference USP at the
exit \textbf{1} of BS ($\hat{B}_{1}(t)$) overlaps with a third
coherent USP ($\hat{A}_{3}(t)$). We consider $R$ and $T$ being
independent on frequency, direction of propagation or polarization
and we treat both reflections symmetrically. In this latest case the
relationship between input and output states on BS reads \cite{Ou}:
\begin{equation}
\left(%
\begin{array}{c}
  \hat{B}_{1}(t) \\
  \hat{B}_{2}(t) \\
\end{array}%
\right)=\left(%
\begin{array}{cc}
 i\sqrt{R} & \sqrt{T}\\
\sqrt{T} &  i\sqrt{R}\\
\end{array}%
\right)\left(%
\begin{array}{c}
 \hat{A}_{1}(t,l_{1}) \\
\hat{A}_{2}(t,l_{2}) \\
\end{array}%
\right),
\end{equation}
where $\hat{B}_{j}(t)$ is the annihilation operator on the output
\textbf{j} of the BS and
$\hat{A}_{j}(t,l_{j})=e^{\hat{O}_{j}(t)}\hat{A}_{j}(t)$ ($j=1,2$).
The Stokes operators at the output \textbf{1} are defined in agree
with the formulas (\ref{S0})-(\ref{S3}) when now the D-matrices are
\begin{equation}
D_{1}(t)=\left(%
\begin{array}{cc}
 \hat{B}_{1}(t) & 0 \\
0 & \hat{A}_{3}(t) \\
\end{array}%
\right),\quad D_{2}(t)=\left(%
\begin{array}{cc}
 0 & \hat{B}_{1}(t) \\
 \hat{A}_{3}(t) & 0 \\
\end{array}%
\right).
\end{equation}
By using the algebra of time-dependent Bose operators developed in
\cite{POP99,POP00}, and summarily exposed in Section \ref{Section1},
for the average values of Stokes parameters on the summary coherent
state $|\alpha(t)\rangle$ we find:
\begin{eqnarray}
\langle\hat{S}_{0,1}(t)\rangle &=&R\bar{n}_{1}(t)+T
\bar{n}_{2}(t)\pm\bar{n}_{3}(t)\nonumber\\
&+&2\sqrt{RT}[\bar{n}_{1}(t)\bar{n}_{2}(t)]^{1/2}e^{-\mu_{1}(t)-\mu_{2}(t)}
\sin{[\Phi_{2}(t)-\Phi_{1}(t)]},\label{interfS0}\\
\langle\hat{S}_{2}(t)\rangle
&=&2\sqrt{T}[\bar{n}_{2}(t)\bar{n}_{3}(t)]^{1/2}e^{-\mu_{2}(t)}\cos{[\varphi_{3}(t)-\Phi_{2}(t)]}\nonumber\\
&+&2\sqrt{R}[\bar{n}_{1}(t)\bar{n}_{3}(t)]^{1/2}e^{-\mu_{1}(t)}\sin{[\varphi_{3}(t)-\Phi_{1}(t)]},\label{interfS1}
\end{eqnarray}
where in (\ref{interfS0}) the sign `$+$' stands for
$\langle\hat{S}_{0}(t)\rangle$ and `$-$' for
$\langle\hat{S}_{1}(t)\rangle$. The average value of $\hat{S}_{3}$
can be easily obtained by shifting (\ref{interfS1}) in phase with
$-\pi/2$. Notice the appearance of the {\it sin}-term in
(\ref{interfS0}) due  particularly to the interference on BS. We
calculate now the correlation functions of Stokes parameters in the
approximation $\gamma_{j}\ll 1$ ($j=1,2$). Below we write down the
results obtained in the frame of the quantum model presented in
Section \ref{Section1};
\begin{eqnarray}
R_{S_{0,1}}(t,t+\tau)&=&\delta(\tau)-h(\tau)\{2\sqrt{RT}[\bar{n}_{1}(t)\bar{n}_{2}(t)]^{1/2}
[R\phi_{1}(t)\pm T\phi_{2}(t)]\cos{[\Phi_{1}(t)-\Phi_{2}(t)]}\nonumber\\
&{}&~~~~~~~~~~~~~~~~+RT[\bar{n}_{1}(t)\phi_{2}(t)-\bar{n}_{2}(t)\phi_{1}(t)]\sin{2[\Phi_{1}(t)-\Phi_{2}(t)]}\}\nonumber\\
&+&g(\tau)\{RT[n_{1}(t)\phi^{2}_{2}(t)+n_{2}(t)\phi^{2}_{1}(t)]\cos^{2}{[\Phi_{1}(t)-\Phi_{2}(t)]}\},\label{corelS01}\\
R_{S_{2}}(t,t+\tau)&=&\!\delta(\tau)\!+h(\tau)n_{3}(t)\{R\phi_{1}(t)\sin{2[\Phi_{1}(t)\!-\!\varphi_{3}(t)]}\!-
\!T\phi_{2}(t)\sin{2[\Phi_{2}(t)\!-\!\varphi_{3}(t)]}\}\nonumber\\
&+&g(\tau)n_{3}(t)\{R\phi^{2}_{1}(t)\cos^{2}{[\Phi_{1}(t)-\varphi_{3}(t)]}+T
\phi^{2}_{2}(t)\sin^{2}{[\Phi_{2}(t)-\varphi_{3}(t)]}\}.\label{corelS2}
\end{eqnarray}
The presence of the additional terms besides $\delta$-function
indicates that the formation of the PS state in all three Stokes
parameters $\hat{S}_{0}$, $\hat{S}_{1}$, $\hat{S}_{2}$ is permitted.
The fluctuation spectra are straightforward with the use of WK
theorem:
\begin{eqnarray}
S_{S_{0,1}}(\Omega,t)&=&1-2L(\Omega)\{2\sqrt{RT}[\bar{n}_{1}(t)\bar{n}_{2}(t)]^{1/2}[R\phi_{1}(t)\pm T\phi_{2}(t)]\cos{[\Phi_{1}(t)-\Phi_{2}(t)]}\nonumber\\
&{}&~~~~~~~~~~~~~~~~+RT[\bar{n}_{1}(t)\phi_{2}(t)-\bar{n}_{2}(t)\phi_{1}(t)]\sin{2[\Phi_{1}(t)-\Phi_{2}(t)]}\}\nonumber\\
&+&4L^{2}(\Omega)\{RT[n_{1}(t)\phi^{2}_{2}(t)+n_{2}(t)\phi^{2}_{1}(t)]\cos^{2}{[\Phi_{1}(t)-\Phi_{2}(t)]}\},\label{aaa}\\
S_{S_{2}}(\Omega,t)&=&1+2L(\Omega)\bar{n}_{3}(t)\{R\phi_{1}(t)\sin{2[\Phi_{1}(t)-\varphi_{3}(t)]}\!-\!T\phi_{2}(t)\sin{2[\Phi_{2}(t)-\varphi_{3}(t)]}\}\nonumber\\
&+&4L^{2}(\Omega)\bar{n}_{3}(t)\{R\phi^{2}_{1}(t)\cos^{2}{[\Phi_{1}(t)-\varphi_{3}(t)]}\!+\!T
\phi^{2}_{2}(t)\sin^{2}{[\Phi_{2}(t)-\varphi_{3}(t)]}\}.\label{spectraS2}
\end{eqnarray}
Since the correlation function of $\hat{S}_{3}$ and its fluctuation
spectrum can be easily obtained by shifting (\ref{corelS2}) and
(\ref{spectraS2}) in phase with $\pi/2$, respectively, their
analytical expression are not presented here. Notice that the above
correlation functions and the fluctuation spectra are determined by
the parameters $R$ and $T$. In what follows we consider the most
common case of a $50/50$ BS, i.e., $R=T=1/2$.

\begin{figure}
\centering
\includegraphics[height=.5\textwidth]{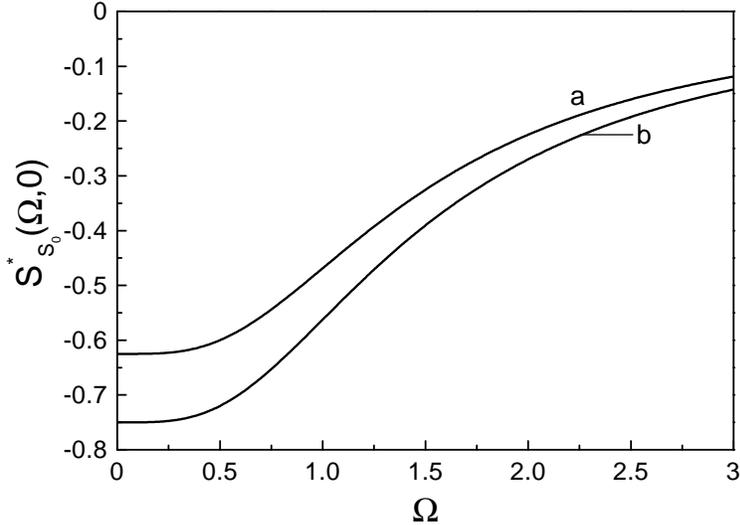}
\caption{Normalized spectral variance $S^{*}_{S_{0}}(\Omega,t)$ for
linear phase difference $\Delta\varphi(t)$ chosen optimal at
$\Omega_{0}=0$. Curves are calculated at time $t=0$, $R=T=0.5$ and
correspond to $\bar{n}_{2,0}=1.5\,\bar{n}_{1,0}$ (a),
$\bar{n}_{2,0}=2\,\bar{n}_{1,0}$ (b).} \label{Fig8}
\end{figure}

\begin{figure}
\centering
\includegraphics[height=.5\textwidth]{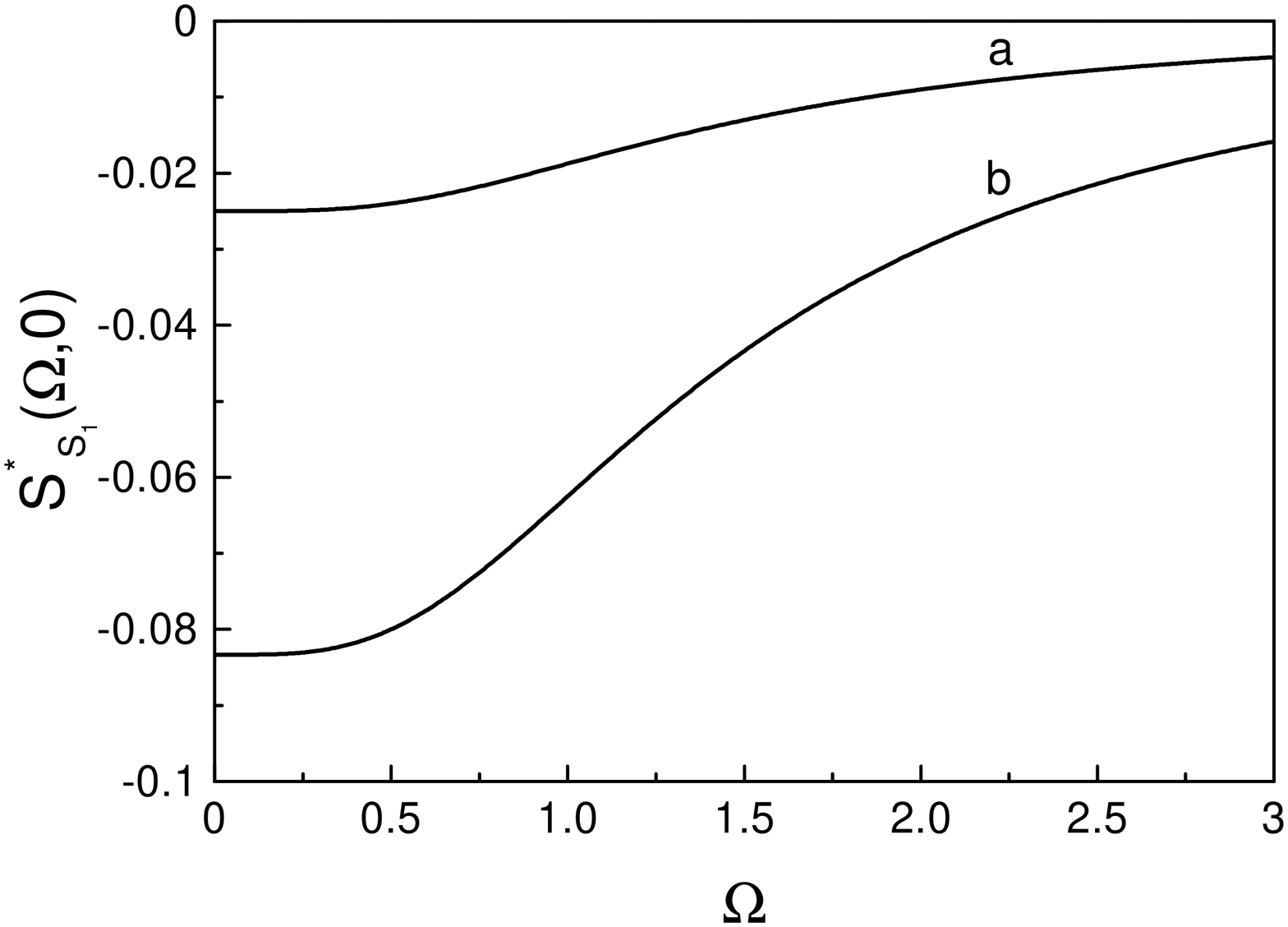}
\caption{Normalized spectral variance $S^{*}_{S_{1}}(\Omega,t)$ for
linear phase difference $\Delta\varphi(t)$ chosen optimal at
$\Omega_{0}=0$. Curves are calculated at time $t=0$, $R=T=0.5$ and
correspond to $\bar{n}_{2,0}=1.5\,\bar{n}_{1,0}$ (a),
$\bar{n}_{2,0}=2\,\bar{n}_{1,0}$ (b).} \label{Fig9}
\end{figure}

\begin{figure}
\centering
\includegraphics[height=.5\textwidth]{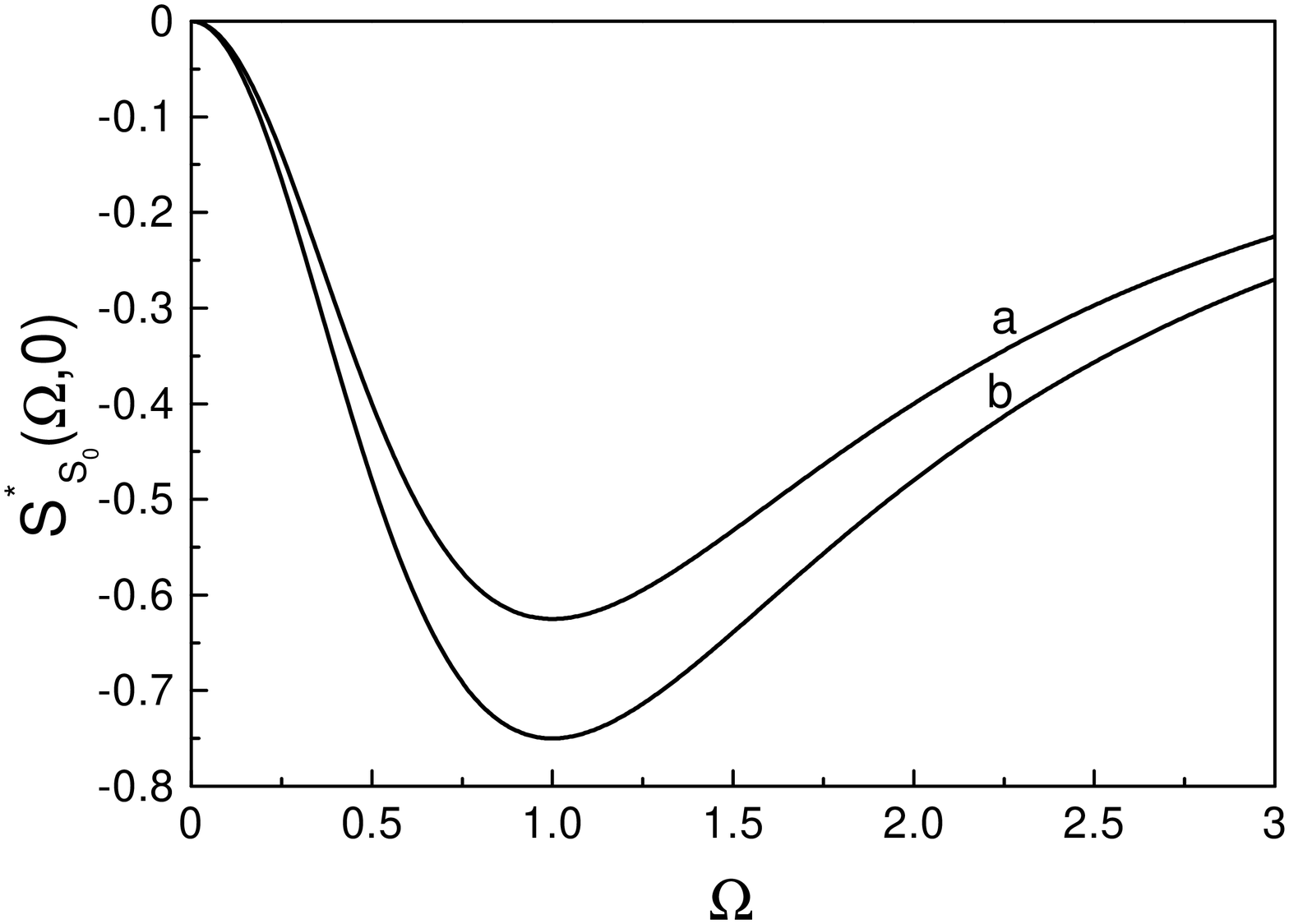}
\caption{As in Fig.\ \ref{Fig8} but for $\Omega_{0}=1$.}
\label{Fig10}
\end{figure}

\begin{figure}
\centering
\includegraphics[height=.5\textwidth]{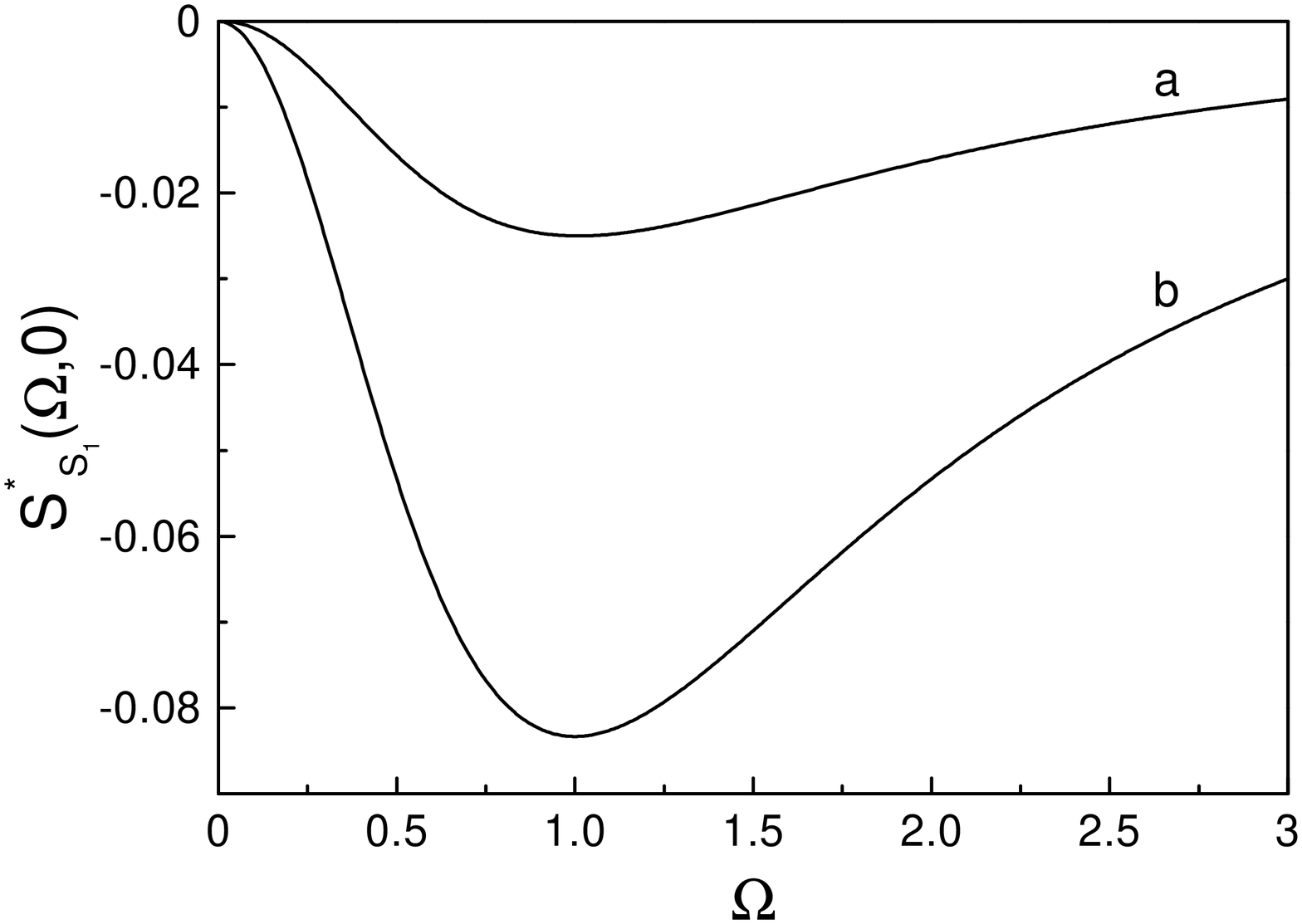}
\caption{As in Fig.\ \ref{Fig9} but for $\Omega_{0}=1$}
\label{Fig11}
\end{figure}

We analyze now the first two Stokes parameters $\hat{S}_{0}$ and
$\hat{S}_{1}$ by addressing the issue concerning the optimization of
the linear phase difference
$\Delta\varphi(t)=\varphi_{1}(t)-\varphi_{2}(t)$. However, due to
the analytical complexity of the expression (\ref{aaa}), we restrict
to the case $\bar{n}_{1}(t)\phi_{2}(t)=\bar{n}_{2}(t)\phi_{1}(t)$
when {\it sin}-term in (\ref{aaa}) vanishes. Thus, the resulting
expression can be minimized if the linear phase difference
$\Delta\varphi(t)$ at a defined frequency $\Omega_{0}$ has the form
\begin{equation}\label{optphase1}
\Delta\varphi^{(\pm)}_{opt}(t)=\arccos{\left[
\frac{R\bar{n}_{1}(t)\pm
T\bar{n}_{2}(t)}{2[\bar{n}_{1}(t)+\bar{n}_{2}(t)]\phi(t)L(\Omega_{0})}
\left(\frac{\bar{n}_{1}(t)}{RT\bar{n}_{2}(t)}\right)^{1/2}\right]}-\phi_{1}(t)+\phi_{2}(t).
\end{equation}
With this adjustment (\ref{optphase1}) for the linear phase
difference between pulses the expressions (\ref{aaa}) achieve the
minimal value
\begin{equation}
S_{S_{0,1}}(\Omega_{0},t)=1-[R\bar{n}_{1}(t)\pm
T\bar{n}_{2}(t)]^{2}\,[\bar{n}_{1}(t)+\bar{n}_{2}(t)]^{-1}.
\end{equation}
Thus, at any frequency $\Omega$ the spectra (\ref{aaa}) take the
form
\begin{equation}\label{arra}
S_{S_{0,1}}(\Omega,t)=S_{S_{0,1}}(\Omega_{0},t)+\frac{[R\bar{n}_{1}(t)\pm
T\bar{n}_{2}(t)]^{2}}{\bar{n}_{1}(t)+\bar{n}_{2}(t)}\frac{[L(\Omega)-L(\Omega_{0})]^{2}}{L^{2}(\Omega_{0})}.
\end{equation}
The fluctuation spectra of $\hat{S}_{0}$ and $\hat{S}_{1}$ for
various relations between pulses intensities, at $\Delta\varphi(t)$
optimized at $\Omega_{0}=0$, are displayed in Fig.\ \ref{Fig8} and
\ref{Fig9}, respectively. We conclude that the adjustment of the
control pulse intensity can be employed to obtain a desired level of
squeezing of quantum fluctuations in $\hat{S}_{0}$ and
$\hat{S}_{1}$. This conclusion remain also valid in case the linear
phase difference is optimized at frequency $\Omega_{0}=1$, the
simultaneous suppression being correspondingly achieved at high
frequencies $\Omega\approx1$ ($\omega\approx1/\tau_{r}$) [Figs.\
\ref{Fig10} and \ref{Fig11}]. With the simplification introduced by
the condition $\bar{n}_{1}(t)\phi_{2}(t)=\bar{n}_{2}(t)\phi_{1}(t)$,
while the suppression achieved in $\hat{S}_{0}$ is large, the one
achieved in $\hat{S}_{1}$ is small.
%%%%%%%%%%%%%%%%%%%%%%%%%%%%%%%%%%%%%%%%%%%%%%%%%%%%%%%%%%%%%%%%%%%%%%%%%%%
%%%%%%%%%%%%%%%%%%%%%%%%%%%%%%%%%%%%%%%%%%%%%%%%%%%%%%%%%%%%%%%%%%%%%%%%%%
%%%%%%%%%%%%%%%%%%%%%%%%%%%%%%%%%%%%%%%%%%%%%%%%%%%%%%%%%%%%%%%%%%%%%%%%%%%
%%%%%%%%%%%%%%%%%%%%%%%%%%%%%%%%%%%%%%%%%%%%%%%%%%%%%%%%%%%%%%%%%%%%%%%%%%
%%%%%%%%%%%%%%%%%%%%%%%%%%%%%%%%%%%%%%%%%%%%%%%%%%%%%%%%%%%%%%%%%%%%%%%%%%%
%%%%%%%%%%%%%%%%%%%%%%%%%%%%%%%%%%%%%%%%%%%%%%%%%%%%%%%%%%%%%%%%%%%%%%%%%%
%%%%%%%%%%%%%%%%%%%%%%%%%%%%%%%%%%%%%%%%%%%%%%%%%%%%%%%%%%%%%%%%%%%%%%%%%%%
%%%%%%%%%%%%%%%%%%%%%%%%%%%%%%%%%%%%%%%%%%%%%%%%%%%%%%%%%%%%%%%%%%%%%%%%%%

However, it is important to remark that in the correlation functions
and fluctuation spectra of $\hat{S}_{0}$ and $\hat{S}_{1}$ there is
no information about the third coherent USP. As a consequence, the
coherent USP can be disregarded in this respect or can be replaced
with coherent vacuum which have an uncertainty identical to that of
the coherent state. Basically, it means that the outgoing
interference USP is already with squeezed quantum fluctuations in
$\hat{S}_{0}$ and $\hat{S}_{1}$ before the overlap with the coherent
pulse. This situation due to the quantum interference is basically
exploited experimentally in \cite{Heersnik} where coherent vacuum is
mixed on a polarization BS with a quadrature-squeezed USP obtained
by using the SPM effect in an optical Kerr fiber ($\chi^{3}$).
Indeed, the outgoing interference USP shows the squeezed quantum
fluctuations of about $-3.7$ and $-3.6$ dB in $\hat{S}_{0}$ and
$\hat{S}_{1}$, respectively. The simultaneous polarization squeezing
of about $-3.4$ dB in both $\hat{S}_{0}$ and $\hat{S}_{1}$ is also
present when two quadrature-squeezed pulses interfere on a
polarization BS \cite{Heersnik,Heersnik2}. A similar squeezing
effect of about $-3$ dB in $\hat{S}_{0}$ and $\hat{S}_{1}$ was
reported in \cite{Bowen1} where two bright quadrature-squeezed
pulses obtained by employing the optical parametrical amplification
($\chi^{2}$) are mixed on a $50/50$ BS.

\begin{figure}
\centering
\includegraphics[height=.5\textwidth]{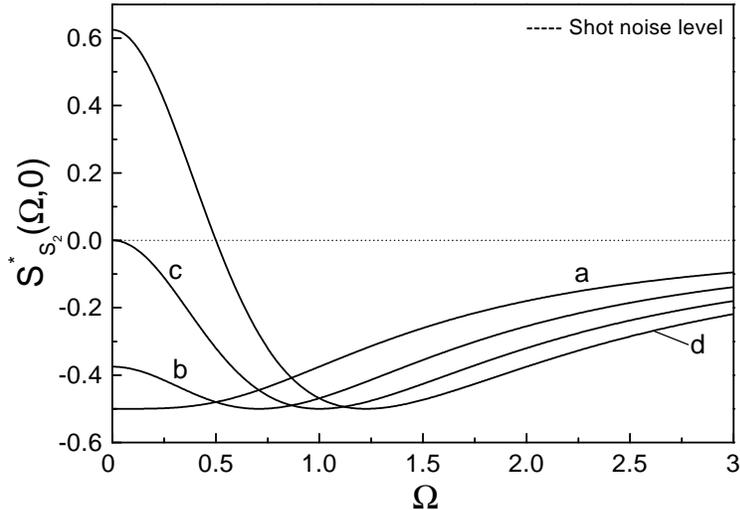}
\caption{Normalized spectral variance $S^{*}_{S_{2}}(\Omega,t)$ for
linear phase difference $\Delta\varphi(t)$ chosen optimal at
$\Omega_{0}=0$. Curves are calculated at time $t=0$, $R=T=0.5$ and
correspond to $\phi_{0}=0.5$ (a), $\phi_{0}=0.75$ (b), $\phi_{0}=1$
(c), $\phi_{0}=1.25$ (d).} \label{Fig12}
\end{figure}

Let us now consider the next two Stokes parameters $\hat{S}_{2}$ and
$\hat{S}_{3}$. We focus on the parameter $\hat{S}_{2}$ since once
the suppression of quantum fluctuations is realized in it, the
parameter $\hat{S}_{3}$ becomes anti-squeezed, and viceversa. The
fluctuation spectrum of Stokes parameter $\hat{S}_{2}$ can be easily
minimized in the case $\phi_{1}(t)=\phi_{2}(t)\equiv\phi(t)$ and
$\varphi_{1}(t)-\varphi_{2}(t)=\pi/2$. Actually, it was noticed
experimentally in \cite{Bowen2} that this relative phase of $\pi/2$
is critical for the detection of squeezing. The linear phase
difference $\Delta\varphi(t)=\varphi(t)-\varphi_{3}(t)$ that at a
defined frequency $\Omega_{0}$ fulfills the condition
$$\cos{2[\phi(t)+\Delta\varphi(t)]}=(R-T)\phi(t)L(\Omega_{0})\sin{2[\phi(t)+\Delta\varphi(t)]}$$
minimizes the expression (\ref{spectraS2}). At this optimal linear
phase difference the fluctuation spectrum becomes
\begin{equation}
S_{S_{2}}(\Omega_{0},t)=1+2\bar{n}_{3}(t)\phi^{2}(t)L^{2}(\Omega_{0})
-2\bar{n}_{3}(t)\phi(t)L(\Omega_{0})[1+(R-T)^2\phi^{2}(t)L^{2}(\Omega_{0})]^{1/2}
\end{equation}
In consequence, at any phase $\Omega$ the normalized spectra is
\begin{eqnarray}
S_{S_{2}}(\Omega,t)&=&S_{S_{2}}(\Omega_{0},t)+2\bar{n}_{3}(t)\phi^{2}(t)[L^{2}(\Omega)-L^{2}(\Omega_{0})]+
2\bar{n}_{3}(t)\phi(t)[L(\Omega)-L(\Omega_{0})]\nonumber\\
&\times&\{1+(R-T)^{2}\phi^{2}(t)L(\Omega_{0})[L(\Omega)+L(\Omega_{0})]\}
[1+(R-T)^{2}\phi^{2}(t)L^{2}(\Omega_{0})]^{-1/2}.
\end{eqnarray}
In Fig.\ \ref{Fig12} we displayed the fluctuation spectrum of
$\hat{S}_{2}$ at time moment $t=0$ for different values of the
nonlinear phase $\phi(t=0)\equiv\phi_{0}$.
%%%%%%%%%%%%%%%%%%%%%%%%%%%%%%%%%%%%%%%%%%%%%%%%%%%%%%%%%%%%%%%%%%%%%%%%%%
%%%%%%%%%%%%%%%%%%%%%%%%%%%%%%%%%%%%%%%%%%%%%%%%%%%%%%%%%%%%%%%%%%%%%%%%%%
The increase of $\phi_{0}$ does not produce a better level of
squeezing but moves the squeezing from low frequencies
$\Omega\approx 0$ to high frequencies $\Omega\approx 1$. At the same
time the squeezing at low frequencies is deteriorating.

Note the obtained suppression of quantum fluctuations due to the
stable spatial overlap of two quadrature-squeezed pulses of about of
$-2.8$ dB in $\hat{S}_{2}$ reported in \cite{Heersnik,Heersnik2}.
Indeed, since the parameter $\hat{S}_{2}$ becomes squeezed, the
parameter $\hat{S}_{3}$ is detected anti-squeezed with about $+23.5$
dB above the shot noise level.

Summarizing, the mixing of two quadrature-squeezed pulses on a
$50/50$ BS followed by the spatial overlap of the outgoing
interference pulse with a coherent field permits the simultaneous
squeezing of quantum fluctuations in the first two Stokes parameters
$\hat{S}_{0}$ and $\hat{S}_{1}$, as well as the squeezing in one of
the remaining two, $\hat{S}_{2}$ or $\hat{S}_{3}$ (here
$\hat{S}_{2}$).
%---------------------------------------------------------------------
\section{Conclusion}\label{Conclusion}
In this work we applied the quantum model developed in \cite{QSO} to
analyze the spatial overlap and interference of quadrature-squeezed
USPs. We showed that the spatial overlap of coherent and
quadrature-squeezed pulses or of two quadrature-squeezed USPs
obtained by employing the SPM effect in the Kerr medium produces the
squeezing in one of the last two Stokes parameters $\hat{S}_{2}$ or
$\hat{S}_{3}$. We revealed that the adjustment of linear phase
difference between pulses leads to the suppression of quantum
fluctuations of $\hat{S}_{2}$ or $\hat{S}_{3}$ at the frequency of
interest. Moreover, we found that the increase of the intensity of
the control USP produces a better level of squeezing.

By investigating the spatial overlap of nonclassical pulses in a
nonlinear anisotropic Kerr medium in the presence of XPM effect we
established that the increase of the intensity of the control pulse
or of one nonlinear coefficient ($\gamma_{2}$) in comparison with
the another one ($\gamma_{1}$) is able to realize the desired level
of squeezing in $\hat{S}_{2}$ or $\hat{S}_{3}$.

We studied the overlap of a coherent pulse with a nonclassical pulse
produced by interfering two quadrature-squeezed USPs on a beam
splitter. By studying the fluctuation spectra of Stokes parameters
we showed that the nonclassical pulse exhibits the simultaneous
squeezing of quantum fluctuations in $\hat{S}_{0}$ and
$\hat{S}_{1}$. Besides, we found that its spatial overlap with a
coherent pulse field produce the squeezing in $\hat{S}_{2}$ or
$\hat{S}_{3}$. \vfill

%%%%%%%%%%%%%%%%%%%%%%%%%%%%%%%%%%%%%%%%%%%%%%%%%%%%%%%%%%%%%%%%%

%---------------------------------------------------------------------
\end{document}